\documentclass[aps, prl, twocolumn, superscriptaddress, groupaddress, floatfix, longbibliography]{revtex4-1}
\usepackage{graphicx}
\usepackage{bm}
\usepackage{amssymb}
\usepackage{color}
\usepackage{amsmath, mathrsfs}
\usepackage{amstext}
\usepackage{float}
\usepackage{braket}
\usepackage{cancel}
\usepackage{oubraces}
\usepackage{latexsym}
\usepackage[usenames,dvipsnames]{xcolor}
\usepackage[colorlinks=true,citecolor=Blue,linkcolor=RubineRed,urlcolor=Blue]{hyperref}
\usepackage{bbold}
\usepackage{bbm}

\usepackage{stackengine}
\usepackage{graphicx}

\DeclareFontFamily{OT1}{pzc}{}
\DeclareFontShape{OT1}{pzc}{m}{it}{<-> s * [0.900] pzcmi7t}{}
\DeclareMathAlphabet{\mathpzc}{OT1}{pzc}{m}{it}

\def\H{\hat{H}}

\def\s{\hat{\sigma}}
\def\sd{\hat{\sigma}^{\dagger}}

\def\dm{\hat{\rho}}
\def\E{\hat{E}}
\usepackage[colorinlistoftodos]{todonotes}

\begin{document}
\title{Addressing the Correlation of Stokes-Shifted Photons Emitted from Two Quantum Emitters}

\author{Adri\'an Juan-Delgado}
\email[]{adrianjuand1996@gmail.com}
\affiliation{Centro de F\'isica de Materiales (CMF-MPC), CSIC-UPV/EHU, 20018, Donostia-San Sebasti\'an, Spain}
\affiliation{Department of Electricity and Electronics, University of the Basque Country (UPV/EHU), Leioa 48940, Spain}

\author{Jean-Baptiste Trebbia}
\affiliation{Universit\'e de Bordeaux, LP2N, F-33405 Talence, France}
\affiliation{Institut d’Optique \& CNRS, LP2N, F-33405 Talence, France}

\author{Ruben Esteban}
\affiliation{Centro de F\'isica de Materiales (CMF-MPC), CSIC-UPV/EHU, 20018, Donostia-San Sebasti\'an, Spain}
\affiliation{Donostia International Physics Center (DIPC), 20018, Donostia-San Sebasti\'an, Spain}

\author{Quentin Deplano}
\affiliation{Universit\'e de Bordeaux, LP2N, F-33405 Talence, France}
\affiliation{Institut d’Optique \& CNRS, LP2N, F-33405 Talence, France}

\author{Philippe Tamarat}
\affiliation{Universit\'e de Bordeaux, LP2N, F-33405 Talence, France}
\affiliation{Institut d’Optique \& CNRS, LP2N, F-33405 Talence, France}

\author{R\'emi Avriller}
\affiliation{Université de Bordeaux, CNRS, LOMA, UMR 5798, F-33400 Talence, France}
\affiliation{Université de Strasbourg, CNRS, Institut de Physique et Chimie des Matériaux de Strasbourg, UMR 7504, F-67000 Strasbourg, France}

\author{Brahim Lounis}
\affiliation{Universit\'e de Bordeaux, LP2N, F-33405 Talence, France}
\affiliation{Institut d’Optique \& CNRS, LP2N, F-33405 Talence, France}

\author{Javier Aizpurua}
\email[]{aizpurua@ehu.eus}
\affiliation{Donostia International Physics Center (DIPC), 20018, Donostia-San Sebasti\'an, Spain}
\affiliation{Ikerbasque, Basque Foundation for Science, 48009 Bilbao, Spain.}
\affiliation{Department of Electricity and Electronics, University of the Basque Country (UPV/EHU), Leioa 48940, Spain}

\date{\today}

\begin{abstract}
In resonance fluorescence excitation experiments, light emitted from solid-state quantum emitters is typically filtered to eliminate the laser photons, ensuring that only red-shifted Stokes photons are detected. However, theoretical analyses of the fluorescence intensity correlation often model emitters as two-level systems, focusing on light emitted exclusively from the purely electronic transition (the zero-phonon line), or they rely on statistical approaches based on conditional probabilities that neglect the quantum coherence between the emitters and the coherence between the electric fields they generate. Here, we propose a model to characterize the correlation of either zero-phonon line photons or Stokes-shifted photons. This model successfully reproduces the experimental correlation of Stokes-shifted photons emitted from two interacting molecules and predicts that this correlation is affected by quantum coherence. 
Besides, we analyze the role of quantum coherence in the Stokes-shifted emission from two distant emitters, showing a sharp peak at zero time delay due to the Hanbury Brown--Twiss effect.
\end{abstract} 

\maketitle

Theoretical descriptions of light emission from quantum emitters typically focus on the case of simple two-level systems (TLSs) \cite{Lehmberg_PRA_I_1970,Agarwal_book_1974, Agarwal_PRA_1977,Griffin_PRA_1982, Ficek_OA_1983, Ficek_PRA_1984, Lawande_OC_1989, Rudolph_1995_PRA, Gillet_PRA_2010, VivasViana_PRR, Downing_PRA_2023, JuanDelgado}. However, many realistic systems--such as solid-state emitters, trapped ions, or trapped atoms--exhibit vibrational degrees of freedom that couple to the electronic states. This coupling gives rise to two main emission channels: the zero-phonon line (ZPL), corresponding to a direct transition between the electronic excited and ground states, and phonon sidebands, which involve transitions to vibrationally excited ground states and appear red-shifted in the emission spectrum. Accurate modeling of these emission processes, particularly in systems involving two coherently interacting quantum emitters, is necessary for the advancement of quantum photonics. For example, these coupled systems can find applications in quantum information \cite{Schafer_Nature_2018, Reina_PRA_2018,Plankensteiner_SR_2015, Facchinetti_PRL_2016,Asenjo_PRX_2017}, quantum-state engineering \cite{Lloyd_Science_1993}, and the generation of entangled photons \cite{JuanDelgado_entanglement_2025}. 

Moreover, in usual resonance fluorescence experiments, the laser is tuned to the ZPL transition, but only the red-shifted (Stokes) photons are detected, minimizing contamination from scattered laser light \cite{Tamarat_JPCA_1999, Moerner_JPCB_2002, Orrit_PRL_1990, Ambrose_Nature_1991}. Under these conditions, the standard TLS framework proves insufficient—particularly for experiments involving emission from interacting solid-state emitters \cite{Hettich_Science_2002,Trebbia_NatComms_2022, Lange_arxiv_2023}. To overcome this problem, the correlation of Stokes-shifted photons can be described using a conditional-probability approach, as introduced in Refs. \cite{Hettich_Science_2002,Trebbia_NatComms_2022}. This approach is based on the calculation of the conditional probabilities of a fluorescence photon being emitted by the system at time $t+\tau$, following an earlier emission at time $t$ that projects the state of the system \cite{Hegerfeldt_PRA_1993,Beige_PRA_1998, Hettich_thesis_2002}. However, this approach does not account for the influence of the quantum coherence between the two emitters nor the quantum coherence of the electric field that they emit (see Supplemental Material \footnote{See Supplemental Material at http://link.aps.org/
supplemental/10.1103/1z52-p73t, which includes Refs. \cite{Stokes_NJP_2018,Trebbia_NatComms_2022,JuanDelgado,Steck_book_2007,Hettich_Science_2002,Hegerfeldt_PRA_1993,Beige_PRA_1998,Hettich_thesis_2002,Glauber_PR_1963,VivasViana_PRR,Jelezko_JCP_1997,Loudon_book_2000}, for the complete expressions of the coherent and incoherent dipole-dipole couplings, the details of all the parameters used in the simulations, the conditional-probability approach to the Stokes-shifted photon correlations, a summary of the typical framework with two-level systems, the complete expression of the correlation of Stokes-shifted photons, details of the experimental setup, further experimental measurements, an analysis of the impact of a larger number of vibrational modes, and the extension of the model for the case of $N$ emitters.}). Thus, this approach can fail to describe experiments where coherence becomes important, as indicated by our results here.  

\begin{figure}[!t] 
	\begin{center}
		\includegraphics[width=0.48\textwidth]{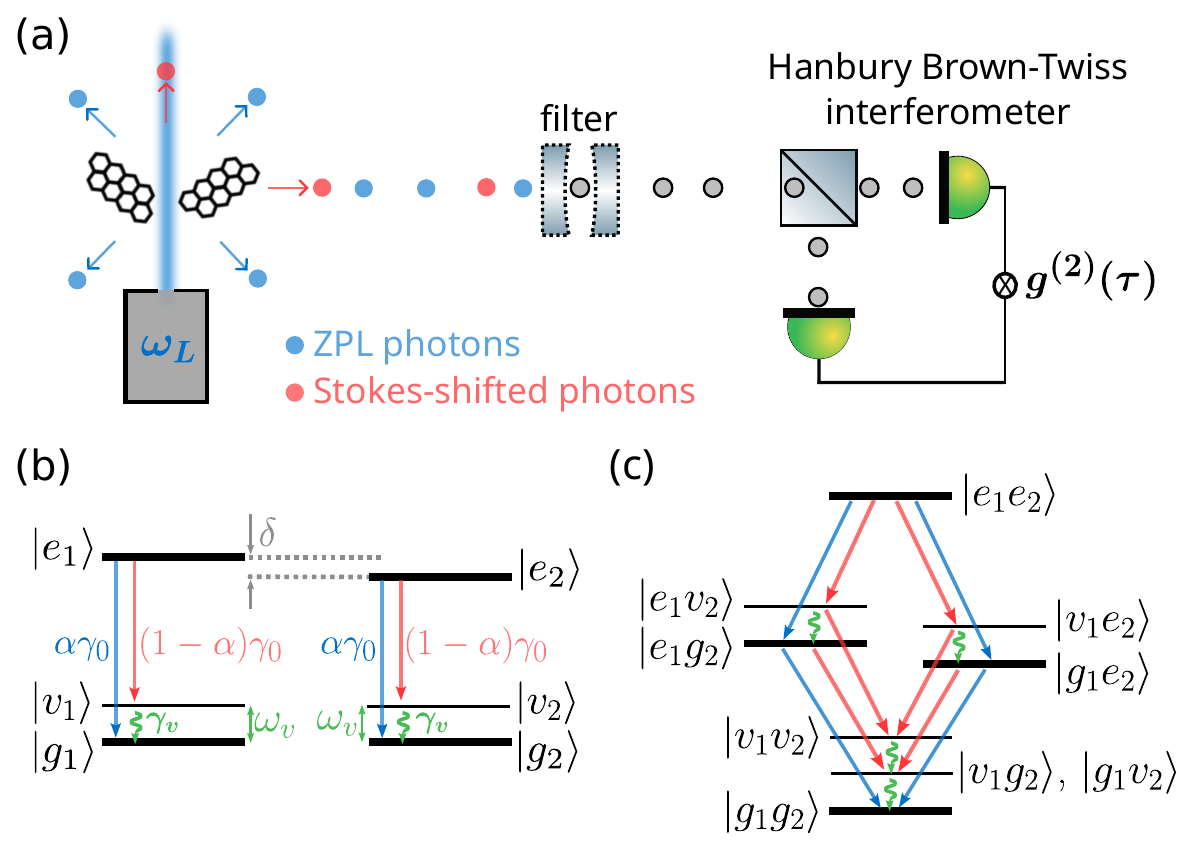}%
		\caption{(a) Schematic representation of light emitted from two interacting emitters, the filtering of this light and the measurement of the intensity correlation $g^{(2)}(\tau)$. A laser beam at frequency $\omega_L$ (in blue) excites resonantly the pure electronic excited state of the quantum emitters, which then can emit a photon at the same frequency (blue circles) or a Stokes-shifted photon (red circles). A filter selects either the ZPL or Stokes-shifted light (gray circles represent these filtered photons). 
       (b,c) Energy levels and relaxation processes of two noninteracting emitters represented in the uncoupled local basis, using the single-emitter representation in (b) and the two-emitter representation in (c). Blue arrows correspond to ZPL transitions, and red arrows to Stokes-shifted transitions.}
        \label{Figure:1}  
	\end{center}
\end{figure}

In this Letter, we present a model that incorporates quantum coherence effects to describe the photon statistics of both zero-phonon line (ZPL) and Stokes-shifted emission from quantum emitters, as illustrated in Fig. \ref{Figure:1}a. We apply this model to analyze the emission from two interacting organic molecules at cryogenic temperature, revealing that the statistics of ZPL and Stokes-shifted photons can differ significantly. 
We further explore the case of two distant, noninteracting emitters. In this regime, the model predicts a sharp feature at zero time delay due to the Hanbury Brown--Twiss effect, and we discuss the conditions necessary for its experimental observation.

We consider two almost identical quantum emitters (indexed by $j=1,2$), with a pure ($0$-phonon) electronic ground state $\ket{g_j}$ and are excited state $\ket{e_j}$. The transition dipole moment between these two states is denoted by $\boldsymbol{\mu}_j$ and the transition frequency by $\omega_j$. The detuning between the two transition frequencies is $\delta = \omega_1 - \omega_2$, with $\delta \ll \omega_1 , \omega_2$. We consider an additional state $\ket{v_j}$ that corresponds to the $1$-phonon state of a vibrational mode of frequency $\omega_v$ in the electronic ground state; see Figs. \ref{Figure:1}b and \ref{Figure:1}c. In the rotating frame at the laser frequency $\omega_L$, the unperturbed Hamiltonian of the system can be written as
\begin{equation}
\H_0 = \hbar \sum_{j=1}^2 \biggr[ \frac{\Delta_j}{2} ( \ket{e_j}\bra{e_j}-\ket{g_j}\bra{g_j}) + \frac{2\omega_v - \Delta_j}{2}\ket{v_j}\bra{v_j} \biggr],
\end{equation}
with $\Delta_j = \omega_j - \omega_L$. Further, the coherent dipole-dipole interaction between the two emitters is described by $\H_{I}=\hbar V(\sd_1 \s_2 + \s_1 \sd_2)$, with $V$ the coupling strength \cite{JuanDelgado,Note1,Stokes_NJP_2018}, and $\s_j=\ket{g_j}\bra{e_j}$ and $\sd_j$ the  ZPL lowering and raising operators of emitter $j$, respectively. This interaction Hamiltonian considers that the emitters couple only through the purely electronic states. The coupling through the $1$-phonon levels is not taken into account because it does not affect the dynamics of the emitters nor their light emission, as we consider short-lived vibrations. 
The total Hamiltonian is, thus, $\H=\H_0 + \H_I + \H_P$, where $\H_{P}=-\frac{\hbar}{2} \sum_{j=1}^2 (\Omega_j \sd_j + \Omega_j^* \s_j)$ is the pumping Hamiltonian. Here, $\Omega_j = \boldsymbol{\mu}_j\cdot \boldsymbol{\mathcal{E}}_j e^{i \boldsymbol{k}_L \cdot \boldsymbol{r}_j}/\hbar$ is the Rabi frequency, with $\boldsymbol{k}_L$ the laser wave vector and $\boldsymbol{\mathcal{E}}_j$ the laser electric field at the position $\boldsymbol{r}_j$ of emitter $j$. 

The state of the interacting emitters can be described by the density matrix $\dm$, whose dynamics is governed by the Markovian master equation,
\begin{align}\label{equation:master_equation_newmodel}
    \frac{d}{dt}\dm &=-\frac{i}{\hbar}[\H, \dm] + \sum_{j=1}^2 \biggl(\frac{\alpha \gamma_{{0}}}{2}\mathcal{D}[\s_j] + \frac{(1-\alpha) \gamma_{{0}}}{2}\mathcal{D}[\ket{v_j}\bra{e_j}] \notag  \\
    &+ \sum_{k\neq j} \frac{\gamma_{jk}}{2} \mathcal{D}[\s_j , \s_k]  + \frac{\gamma_{v}}{2}\mathcal{D}[\ket{g_j}\bra{v_j}] \biggr)\rho , 
\end{align}
where we have introduced the dissipators $\mathcal{D}[\hat{A},\hat{B}]\dm=2\hat{A}\dm\hat{B}^{\dagger}-\hat{B}^{\dagger}\hat{A}\dm-\dm \hat{B}^{\dagger}\hat{A}$ and $\mathcal{D}[\hat{A}]=\mathcal{D}[\hat{A},\hat{A}]$ \cite{Breuer_Petruccione_book}. Here, $\alpha$ is the combined Debye-Waller/Franck-Condon factor, given by the fraction of photons emitted in the ZPL line \cite{Trebbia_NatComms_2022, JuanDelgado,Basche_book,Note1}, and $\gamma_{{0}}$ is the total decay rate from the excited state $\ket{e_j}$. Thus, we have fixed $\alpha \gamma_0$ as the decay rate in the ZPL and $(1-\alpha)\gamma_{{0}}$ as the decay rate into the $1$-phonon state. $\gamma_{jk}$ is the crossed-decay rate including the effect of $\alpha$ \cite{Note1}, and $\gamma_v$ is the vibrational decay rate, which we consider to be much larger than $V$.

\begin{figure*}[!t] 
	\begin{center}
		\includegraphics[width=0.98\textwidth]{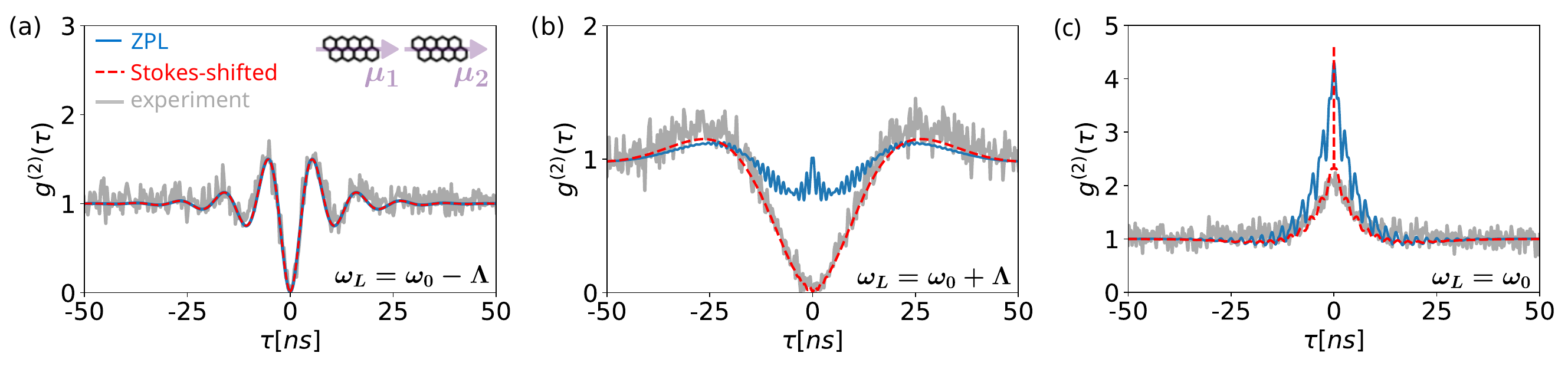}%
		\caption{Comparison of the correlation of ZPL photons and Stokes-shifted photons emitted from two strongly interacting dibenzanthanthrene (DBATT) molecules. The molecules are in a J-aggregate configuration, as depicted in the inset in (a), and have $1/\gamma_0 =7.4$ ns. The laser is tuned resonantly to the (a) superradiant state $\ket{\Lambda_-}$ ($\omega_L = \omega_0 - \Lambda$), (b) subradiant state $\ket{\Lambda_+}$ ($\omega_L = \omega_0 + \Lambda$) and (c) two-photon resonance ($\omega_L = \omega_0$). The simulated intensity correlation $g^{(2)}(\tau)$ of (solid blue line) ZPL and (dashed red line) Stokes-shifted light are plotted as a function of time delay $\tau$. Solid gray line corresponds to the experimental results reported in Ref. \cite{Trebbia_NatComms_2022}. All the parameters are specified in Supplemental Material \cite{Note1}.}  
        \label{Figure:2}  
	\end{center}
\end{figure*}

Assuming that the transition dipole moments of both emitters are identical ($\boldsymbol{\mu}_1 = \boldsymbol{\mu}_2$), the intensity correlation is given by \cite{Mandel_book_1995, Loudon_book_2000,Gerry_book_2005,Steck_book_2007}
\begin{equation} \label{equation:g2_general_equation}
    g^{(2)}(\tau)=\frac{\braket{\E^{(-)}(0)\E^{(-)}(\tau)\E^{(+)}(\tau)\E^{(+)}(0)}_{ss}}{\braket{E^{(-)}(0)E^{(+)}(0)}_{ss}^2 },
\end{equation}
where $\E^{(+)}(\tau)$ and $\E^{(-)}(\tau)$ are the positive-frequency and negative-frequency electric field operators in the Heinsenberg picture, and $\braket{\hat{A}}_{ss}=\text{Tr}(\hat{A}\dm_{ss})$ is the expected value of $\hat{A}$ at the steady state $\dm_{ss}$. The experimental filtering is introduced through the proper definition of the electric field operators. The correlation of ZPL photons can be calculated using  $\hat{E}_{\text{ZPL}}^{(+)}(t) /\xi_{\text{ZPL}} =  \s_1 (t) + e^{i\phi_{\text{ZPL}}} \s_2 (t)$ in Eq.~(\ref{equation:g2_general_equation}), whereas the correlation of Stokes-shifted photons can be obtained through $\hat{E}_{\text{St}}^{(+)}(t) /\xi_{\text{St}}= \ket{v_1}\bra{e_1} (t) + e^{i\phi_{\text{St}}}\ket{v_2}\bra{e_2} (t) $. 
Here, $\xi_{\text{ZPL}}$ and $\xi_{\text{St}}$ are proportionality constants that do not affect the intensity correlation and depend on the direction of the dipoles, the direction of detection $\hat{\boldsymbol{k}}_d$ and the mean frequency of emission [i.e., $\omega_0 = (\omega_1 + \omega_2 )/2$ in the case of  $\xi_{\text{ZPL}}$ and $\omega_0 - \omega_v$ in the case of $\xi_{\text{St}}$] \cite{Lehmberg_PRA_I_1970,Ficek_PRA_1984, Rudolph_1995_PRA}. Additionally, $\phi_{\text{ZPL}}=- (n \omega_0 /c) \hat{\boldsymbol{k}}_d  \cdot \boldsymbol{r}_{12}$ and $\phi_{\text{St}}=-[(\omega_0 - \omega_v) /c]\hat{\boldsymbol{k}}_d n  \cdot \boldsymbol{r}_{12}$ \cite{Skornia_PRA_2001, Agarwal_book_1974}, with $n$ the refractive index of the host medium, $c$ the speed of light in vacuum, and $\boldsymbol{r}_{12} = \boldsymbol{r}_{2} - \boldsymbol{r}_{1}$.

We demonstrate next that the correlation of ZPL photons and of Stokes-shifted photons can be drastically different. With this purpose, we simulate the experimental configuration in Ref. \cite{Trebbia_NatComms_2022}, which measures the statistics of the Stokes-shifted photons emitted from two strongly interacting dibenzanthanthrene (DBATT) molecules (in a J-aggregate configuration \cite{Note1}) in the normal direction to $\boldsymbol{\mu}_j$ and $\boldsymbol{r}_{12}$ (i.e., $\phi_{\text{ZPL}}=\phi_{\text{St}}=0$). The mean frequency $\omega_0$ of these molecules corresponds to a vacuum wavelength of $618$ nm. For simulations, we consider the DBATT vibrational mode $\hbar\omega_v = 31.86$ meV \cite{Jelezko_JCP_1997}, and $1/\gamma_v = 10$ ps based on experiments in Ref. \cite{Zirkelbach_JCP_2022}. At $\tau \gg 1/\gamma_v$, the simulations are not affected by the value of $\omega_v$ and $\gamma_v$ nor by including a larger number of vibrational modes (see Supplemental Material \cite{Note1}).
Other molecular parameters (including $V$, $\gamma_{12}$ and $\delta$) are extracted from independent measurements of the excitation spectra \cite{Note1}, so that no fitting parameters are considered.

We plot in Fig. \ref{Figure:2}a the intensity correlation $g^{(2)}(\tau)$ when the laser is tuned resonantly to the superradiant state, which in the J-aggregate configuration corresponds to the hybrid state $\ket{\Lambda_-}=-\sin\theta \ket{g_1 e_2}+\cos\theta\ket{e_1 g_2}$ \cite{JuanDelgado}, with transition frequency $\omega_0 - \Lambda$. Here, $\tan(2\theta)=-2V/\delta$ and $\Lambda = \sqrt{V^2 + (\delta/2)^2}$. We find that the intensity correlation of Stokes-shifted photons (dashed red line) is almost identical to the correlation of ZPL photons (solid blue line), both exhibiting antibunching and Rabi oscillations. This effective TLS behavior arises because the excitation is sufficiently weak and the laser is far detuned from both the two-photon transition and the transition to the antisymmetric state, thereby isolating the dynamics of the superradiant state \cite{JuanDelgado}. The intensity correlation obtained experimentally in Ref. \cite{Trebbia_NatComms_2022} (solid gray line) is well reproduced by the two simulations. On the other hand, Fig. \ref{Figure:2}b shows the intensity correlation when the laser is tuned resonantly to the subradiant state $\ket{\Lambda_+}=\cos\theta \ket{g_1 e_2}+\sin \theta\ket{e_1 g_2}$, with transition frequency $\omega_0 + \Lambda$. In this regime, the simulated correlations for Stokes-shifted and ZPL photons are significantly different. The Stokes-shifted photons again show antibunching, Rabi oscillations, and excellent agreement with the experiment. In contrast, the ZPL photon correlation also displays clear Rabi oscillations, but with \( g^{(2)}(0) \approx 1 \) and an additional faster oscillation at frequency \( 2\Lambda \), indicating interference between the superradiant and subradiant pathways~\cite{JuanDelgado}. Such interference does not affect the Stokes-shifted photon correlation, since the emitters couple only through the ZPL (see Supplemental Material \cite{Note1}).

Further, we show in Fig. \ref{Figure:2}c the intensity correlation when the laser is tuned to the two-photon resonance ($\omega_L = \omega_0$) corresponding to half the frequency between $\ket{g_1 g_2}$ and $\ket{e_1 e_2}$. In this case, both the correlations of ZPL photons and of Stokes-shifted photons are bunched, as this laser frequency enables the resonant excitation of the doubly excited state $\ket{e_1 e_2}$, strongly increasing the probability of emitting photons in cascade \cite{Beige_PRA_1998, JuanDelgado}. Additionally, the two correlations exhibit oscillations of frequency $\Lambda$, corresponding to the detuning between the laser frequency and the transition frequency of the superradiant state. However, the ZPL correlation exhibits more pronounced oscillations and does not capture well the experimental measurements, whereas the Stokes-shifted correlation reproduces them very well. These results reveal that the correlation of Stokes-shifted photons and of ZPL photons emitted from two strongly interacting quantum emitters can be very different, and they emphasize the importance of an accurate description of each experimental configuration.

\begin{figure}[!b] 
	\begin{center}
		\includegraphics[width=0.483\textwidth]{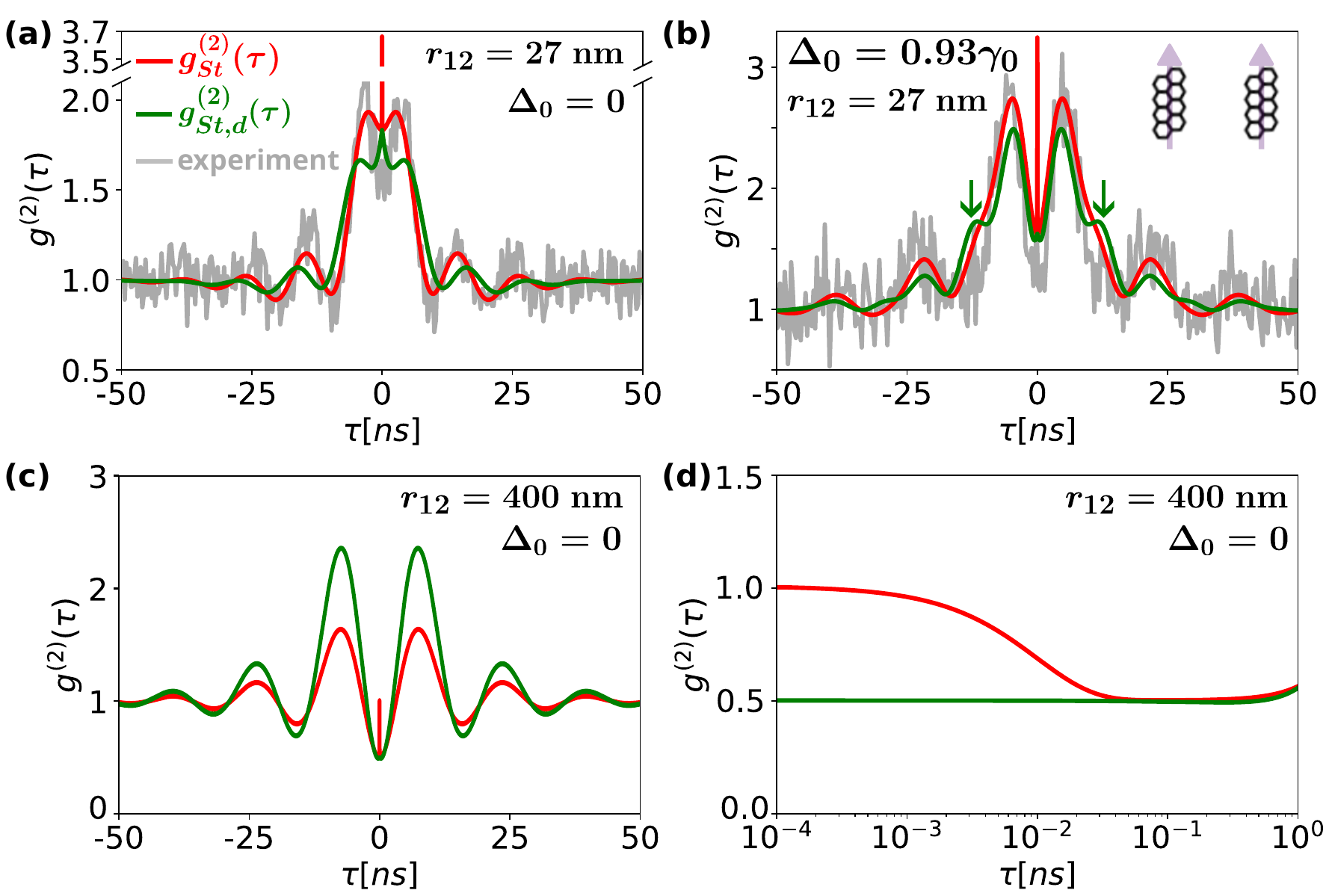}%
		\caption{Impact of quantum coherence in the correlation of Stokes-shifted photons emitted from two DBATT molecules. The molecules are in an H-aggregate configuration, as depicted in the inset in (b), and have $1/\gamma_0 =7.4$ ns. We consider $r_{12}=27$ nm in (a, b) and $r_{12}=400$ nm in (c, d). The laser is tuned to the two-photon resonance $\Delta_0 = 0$ in all panels (with $\Delta_0 = \omega_0 - \omega_L$), except in (b) where $\Delta_0 = 0.93 \gamma_0$. Red lines correspond to the simulation using the full model including quantum coherence in the emission, whereas green lines correspond to the simulation neglecting this coherence. Gray lines in (a,b) correspond to experimental measurements. The rest of the parameters are specified in Supplemental Material \cite{Note1}. }  
        \label{Figure:3}  
	\end{center}
\end{figure}

Next, we investigate the role of quantum coherence in the correlation of Stokes-shifted photons emitted from two quantum emitters. The off-diagonal elements of the electric field intensity operator expressed in the uncoupled basis can be associated with the first-order coherence of the electric field emitted \cite{Glauber_PR_1963,Note1}, and the coherence between the emitters is encoded in the off-diagonal elements of the density matrix \cite{Cohen_book1_1986}. For this analysis, we first substitute $\hat{E}_{\text{St}}^{(+)}(t) / \xi_{\text{St}}= \ket{v_1}\bra{e_1} (t) + \ket{v_2}\bra{e_2} (t)$ (where $\phi_{\text{St}}=0$) into the general expression of the intensity correlation in Eq. (\ref{equation:g2_general_equation}). After some algebra (see Supplemental Material \cite{Note1}), we obtain that the Stokes-shifted correlation is the sum of three contributions
\begin{equation} \label{Eq:g2_Stokes}
    g_{\text{St}}^{(2)}(\tau) = \frac{G_{d}^{(2)}(\tau)}{\braket{\hat{I}_{\text{St}}(0)}_{ss}^2} + \frac{G_{coh, I}^{(2)}(\tau)}{\braket{\hat{I}_{\text{St}}(0)}_{ss}^2} + \frac{G_{coh, \rho}^{(2)}(\tau)}{\braket{\hat{I}_{\text{St}}(0)}_{ss}^2}  .
\end{equation}
We first consider the denominator, where we have introduced the operator $\hat{I}_{\text{St}}(\tau)= \hat{E}_{\text{St}}^{(-)}(\tau)\hat{E}_{\text{St}}^{(+)}(\tau)$ (which is proportional to the fluorescence intensity), with the steady-state expected value
\begin{equation} \label{Eq:mean_value_intensity_Stokes}
    \frac{\braket{\hat{I}_{\text{St}}(0)}_{ss}}{|\xi_{\text{St}}|^2}= 2 p_{ee} + p_{eg} + p_{ge} + p_{ev} +p_{ve} + 2\text{Re}\rho_{ev,ve} .
\end{equation}
Here, $p_{ab}=\braket{a_1 b_2|\dm_{ss}| a_1 b_2}$ is the population of the state $\ket{a_1 b_2}$, 
and $\rho_{ev,ve}=\braket{e_1 v_2|\dm_{ss} | v_1 e_2}$ is an off-diagonal element of the density matrix, which vanishes as long as the emitters couple with each other only through the pure electronic states. The use of the uncoupled local basis $\ket{a_1 b_2}$ (with $a_j ,b_j \in\{e_j ,g_j ,v_j \}$) is convenient because the emission of Stokes-shifted photons projects the system into a localized state of this basis \cite{Hettich_thesis_2002}. 

Further, we have decomposed the correlation in Eq. (\ref{Eq:g2_Stokes}) into three different contributions. The first contribution involves only diagonal elements of $\dm_{ss}$ and $\hat{I}_{\text{St}}(\tau)$, and its numerator is given by
\begin{equation}
\begin{split}
    &\frac{G_{d}^{(2)}(\tau)}{|\xi_{\text{St}}|^4}= p_{ee} [ \bra{v_1 e_2}\hat{I}_{\text{St}}(\tau)\ket{v_1 e_2} + \bra{e_1 v_2}\hat{I}_{\text{St}}(\tau)\ket{e_1 v_2} ]  \\
    &\:\:\:\:\:\:\:\:\: + p_{eg} \bra{v_1 g_2}\hat{I}_{\text{St}}(\tau)\ket{v_1 g_2} + p_{ge} \bra{g_1 v_2}\hat{I}_{\text{St}}(\tau)\ket{g_1 v_2} \\
    &\:\:\:\:\:\:\:\:\: + p_{ev} \bra{v_1 v_2} \hat{I}_{\text{St}}(\tau) \ket{v_1 v_2} + p_{ve} \bra{v_1 v_2} \hat{I}_{\text{St}}(\tau) \ket{v_1 v_2}  
\end{split}   
\end{equation}
The different terms in this expression are related to the six different decay paths leading to the emission of Stokes-shifted photons (red arrows in Fig. \ref{Figure:1}c). The contribution $g_{\text{St,d}}^{(2)}(\tau) = G_{d}^{(2)}(\tau)/\braket{\hat{I}_{\text{St}}(0)}_{ss}^2$ to the intensity correlation in Eq. (\ref{Eq:g2_Stokes}) is independent of quantum coherence and it can be approximated by the correlation obtained with the conditional-probability approach \cite{Note1}. Moreover, 
\begin{equation}
    \frac{G_{coh, I}^{(2)}(\tau)}{|\xi_{\text{St}}|^4} = p_{ee} [\bra{v_1 e_2}\hat{I}_{\text{St}}(\tau)\ket{e_1 v_2} + \bra{e_1 v_2}\hat{I}_{\text{St}}(\tau)\ket{v_1 e_2}] 
\end{equation}
is proportional to the population $p_{ee}$ and to the off-diagonal elements $\bra{v_1 e_2}\hat{I}_{\text{St}}(\tau)\ket{e_1 v_2}$ and $\bra{e_1 v_2}\hat{I}_{\text{St}}(\tau)\ket{v_1 e_2}$ of the intensity operator $\hat{I}_{\text{St}}(\tau)$. ${G_{coh, I}^{(2)}(\tau)}$ can be interpreted as the quantum interference between the emission paths of the two molecules (i.e., the Hanbury Brown--Twiss effect \cite{HBT_Nature_1956,Fano_AJP_1961,Aspect_arxiv_2020}). Last, $G_{coh, \rho}^{(2)} (\tau)$ includes all terms involving off-diagonal elements of the density matrix \cite{Note1} and, thus, the coherence between the emitters. 

We first analyze the role of quantum coherence in the case of two DBATT molecules separated by a short distance ($r_{12}=27$ nm). To this aim, we compare new experimental measurements of the correlation of Stokes-shifted photons emitted from a molecular H-aggregate configuration (the experimental setup in Supplemental Material \cite{Note1}) with the simulated intensity correlations obtained with the complete model [Eq. (\ref{Eq:g2_Stokes})], as well as with the correlation $g_{\text{St,d}}^{(2)}(\tau)= G_{d}^{(2)}(\tau)/\braket{\hat{I}_{\text{St}}(0)}_{ss}^2$, obtained by neglecting the role of quantum coherence in
the emission. In particular, we plot in Figs. \ref{Figure:3}a and \ref{Figure:3}b the correlation of Stokes-shifted photons for this molecular pair when the laser is tuned to $\omega_L = \omega_0$ (resonantly to the two-photon resonance) and to $\omega_L = \omega_0 - 0.93 \gamma_0$ (slightly detuned from the two-photon resonance), respectively. The rest of the parameters are specified in Supplemental Material \cite{Note1}. The simulated $g_{\text{St}}^{(2)}(\tau)$ (red lines) shows a good agreement with the experimental measurements (gray lines) and appreciable differences from $g_{\text{St,d}}^{(2)}(\tau)$ (green line). Specifically, if coherences are neglected in the emission, we observe that (i) the amplitude of the oscillations is notably modified, and (ii) a bump emerges at $\tau\approx \pm10$ ns in Fig. \ref{Figure:3}b (see the green arrows), which is not measured in experiments. These two differences are due to $G_{coh,\rho}^{(2)}(\tau)$. 
Furthermore, $g_{\text{St}}^{(2)}(\tau)$ exhibits an extremely narrow peak at $\tau=0$, with width comparable to the vibrational lifetime, which is due to $G_{coh,I}^{(2)}(\tau)$ and is not resolved experimentally \cite{Note1}. In Supplemental Material \cite{Note1}, we provide additional measurements from different molecular pairs, which also provide good agreement with the simulations obtained with the complete model.

Finally, we analyze the correlation of Stokes-shifted photons emitted from two distant emitters and find that quantum coherence again plays an important role. We plot in Figs. \ref{Figure:3}c and \ref{Figure:3}d the simulated $g_{\text{St}}^{(2)}(\tau)$ (red lines) and $g_{\text{St,d}}^{(2)}(\tau)$ (green lines) for two detuned emitters ($\delta=5\gamma_0$) separated by $r_{12}=400$ nm (which yields a negligible dipole-dipole coupling), and $\omega_L = \omega_0$. The rest of the parameters are specified in Supplemental Material \cite{Note1}. Figure \ref{Figure:3}c shows that both simulations exhibit oscillations at the generalized Rabi frequency \cite{Steck_book_2007}, which at this laser frequency is equal to $\sqrt{(\delta/2)^2 + |\Omega|^2}$. The amplitude of these oscillations is significantly affected by quantum coherence, specifically by $G_{coh,\rho}^{(2)}(\tau)$. Further, the complete model yields $g_{\text{St}}^{(2)}(\tau=0)=1$, with a fast decay to $0.5$ in the timescale of the vibrational lifetime (tens of picoseconds, see Fig. \ref{Figure:3}d), which is a consequence of the Hanbury Brown--Twiss effect between the Stokes-shifted light emitted from the two molecules \cite{Note1}. This fast decay is due to the loss of the initial coherence encoded in $G_{coh,I}^{(2)}(\tau)$ and is attributed to the influence of the internal vibrations of the emitters acting as dephasing channel. 
This behavior of the Stokes-shifted correlation from two distant emitters is analogous to that of the ZPL correlation from two uncorrelated, equally pumped emitters \footnote{We note that the ZPL correlation from two distant, equally pumped emitters can be different from $1$ at delay $\tau=0$, depending on the driving strength and the detection direction \cite{Wolf_PRL_2020, Singh_PRL_2025, Bojer_arXiv_2024}. However, when the emitters are uncorrelated, as in the case of strong driving or large dephasing, the ZPL correlation becomes equal to $1$ at $\tau=0$.}, as it occurs in the limit of strong driving or large dephasing \cite{Auffeves_NJP_2011}, for which available experiments have reported, as far as we know, $g^{(2)}(\tau=0)\approx0.5$ \cite{Diedrich_PRL_1987,Itano_PRA_1988,Gomer_PRA_1998,Pezzagna_NJP_2010}, in contrast with theoretical analysis that predicts a value of $1$ \cite{Auffeves_NJP_2011, Meiser_PRA_2010, Jones_JPB_2016, Wolf_PRL_2020, Singh_PRL_2025, Bojer_arXiv_2024}. Our analysis thus clarifies how this discrepancy is due to an insufficient time resolution of the detectors in the experiments.  

In summary, we have presented a model to address the correlation of Stokes-shifted photons emitted from quantum emitters, as well as that of the ZPL photons. We have shown an excellent agreement between the simulations obtained using this model and experimental measurements of the correlation of the Stokes-shifted photons emitted from two interacting DBATT molecules (using experimental data from Ref. \cite{Trebbia_NatComms_2022}, as well as new experimental measurements). Additionally, we have revealed that the intensity correlation of ZPL photons can exhibit significant differences with respect to the intensity correlation of Stokes-shifted photons depending on the molecular and laser parameters. Furthermore, we have provided evidence that quantum coherence can impact the emission of Stokes-shifted photons when the emitters are interacting and also when they do not interact. Finally, we have found that detectors with time resolutions smaller than the lifetime of the vibrations are needed to measure $g^{(2)}(\tau=0)=1$ in experiments on Stokes-shifted emission from two resonantly driven distant emitters.

Therefore, this work provides a foundation for extending the study of photon correlations emitted by systems with interacting quantum emitters, particularly in platforms that involve both electronic and vibrational states. Examples include atoms and ions in optical, magnetic, or electric traps, as well as solid-state systems like quantum dots coupled to phonons and defect centers in diamonds, carbon nanotubes, and two-dimensional materials. This framework can also be generalized to ensembles with more than two coupled emitters \cite{Note1}, enabling the exploration of collective quantum phenomena in larger emitter networks.


\begin{acknowledgments}
\textit{Acknowledgments}.-- We thank Alejandro Gonzalez-Tudela, \'Alvaro Nodar and Jorge Olmos-Trigo for insightful discussions. A.J.D., R.E., and J.A. acknowledge financial support through Grant No. PID2022-139579NB-I00 funded by MICIU/AEI/10.13039/501100011033 and by ERDF/EU, through Grant No. IT 1526-22 funded by the Department of Science, Universities and Innovation of the Basque Government; and also through the CSIC Research Platform PTI-001 (Project No. 20219PT023). A.J.D. acknowledges financial support through the Grant No. PRE2020-095013 funded by MICIU/AEI/10.13039/501100011033 and by ``ESF Investing in your future". J.-B.T., Q.D., P.T. and B.L. acknowledge the financial support from the French National Agency for Research (Grant No. ANR-22-CE47-0015), Région Nouvelle-Aquitaine, Idex Bordeaux (Research Program GPR Light), and the EUR Light S\&T (PIA3 Program, Grant No. ANR-17-EURE-0027). R.A. acknowledges financial support by EUR Light S$\&$T Graduate Program (PIA3 Program “Investment for the Future,” Grant No. ANR-17-EURE-0027), IdEx of the University of Bordeaux / Grand Research Program GPR LIGHT, Euskampus Transnational Common Laboratory QuantumChemPhys, and Quantum Matter Bordeaux. We acknowledge the Laboratory for Transborder Cooperation LTC TRANS-LIGHT from the University of Bordeaux and the University of the Basque Country.  \\

\textit{Data availability}.-- The data that support the findings of this article are openly available\cite{data_availability}.
\end{acknowledgments}

\bibliographystyle{apsrev4-1}
\bibliography{bibliography} 
\end{document}


\title{Supplemental Material: Addressing the correlation of Stokes-shifted photons emitted from two quantum emitters}

\author{Adri\'an Juan-Delgado}
\email[]{adrianjuand1996@gmail.com}
\affiliation{Centro de Física de Materiales (CMF-MPC), CSIC-UPV/EHU, 20018, Donostia-San Sebasti\'an, Spain}
\affiliation{Department of Electricity and Electronics, University of the Basque Country (UPV/EHU), Leioa 48940, Spain}

\author{Jean-Baptiste Trebbia}
\affiliation{Universit\'e de Bordeaux, LP2N, F-33405 Talence, France}
\affiliation{Institut d’Optique \& CNRS, LP2N, F-33405 Talence, France}

\author{Ruben Esteban}
\affiliation{Centro de Física de Materiales (CMF-MPC), CSIC-UPV/EHU, 20018, Donostia-San Sebasti\'an, Spain}
\affiliation{Donostia International Physics Center (DIPC), 20018, Donostia-San Sebasti\'an, Spain}

\author{Quentin Deplano}
\affiliation{Universit\'e de Bordeaux, LP2N, F-33405 Talence, France}
\affiliation{Institut d’Optique \& CNRS, LP2N, F-33405 Talence, France}

\author{Philippe Tamarat}
\affiliation{Universit\'e de Bordeaux, LP2N, F-33405 Talence, France}
\affiliation{Institut d’Optique \& CNRS, LP2N, F-33405 Talence, France}

\author{R\'emi Avriller}
\affiliation{Université de Bordeaux, CNRS, LOMA, UMR 5798, F-33400 Talence, France}
\affiliation{Université de Strasbourg, CNRS, Institut de Physique et Chimie des Matériaux de Strasbourg, UMR 7504, F-67000 Strasbourg, France}

\author{Brahim Lounis}
\affiliation{Universit\'e de Bordeaux, LP2N, F-33405 Talence, France}
\affiliation{Institut d’Optique \& CNRS, LP2N, F-33405 Talence, France}

\author{Javier Aizpurua}
\email[]{aizpurua@ehu.eus}
\affiliation{Donostia International Physics Center (DIPC), 20018, Donostia-San Sebasti\'an, Spain}
\affiliation{Ikerbasque, Basque Foundation for Science, 48009 Bilbao, Spain.}
\affiliation{Department of Electricity and Electronics, University of the Basque Country (UPV/EHU), Leioa 48940, Spain}

\date{\today}

\maketitle

\onecolumngrid
\tableofcontents
\section{SI. Expressions of the coherent and incoherent dipole-dipole couplings} \label{Appendix:couplings_configurations}
In this section we provide the expressions of the coherent dipole-dipole coupling $V$ and the dissipative dipole-dipole coupling $\gamma_{12}$. These expressions are
\begin{subequations} \label{EqSM:complete_expressions_coupling}
    \begin{align}
        V &= \frac{3}{4} \alpha \gamma_0 \biggr[- \biggr(\hat{ \boldsymbol{\mu}}_1 \cdot\hat{ \boldsymbol{\mu}}_2 - (\hat{ \boldsymbol{\mu}}_1 \cdot \hat{\boldsymbol{r}}_{12})(\hat{ \boldsymbol{\mu}}_2 \cdot \hat{\boldsymbol{r}}_{12}) \biggr)\frac{\cos(k r_{12})}{k r_{12}} + \biggr(\hat{ \boldsymbol{\mu}}_1 \cdot \hat{ \boldsymbol{\mu}}_2 - 3(\hat{ \boldsymbol{\mu}}_1 \cdot \hat{\boldsymbol{r}}_{12})(\hat{ \boldsymbol{\mu}}_2 \cdot \hat{\boldsymbol{r}}_{12}) \biggr)\biggr(\frac{\sin(k r_{12})}{(k r_{12})^2} + \frac{\cos(k r_{12})}{(k r_{12})^3}\biggr) \biggr],\\
        \gamma_{12} &= \frac{3}{2} \alpha \gamma_0 \biggr[ \biggr(\hat{ \boldsymbol{\mu}}_1 \cdot \hat{ \boldsymbol{\mu}}_2 - (\hat{ \boldsymbol{\mu}}_1 \cdot \hat{\boldsymbol{r}}_{12})(\hat{ \boldsymbol{\mu}}_2 \cdot \hat{\boldsymbol{r}}_{12}) \biggr)\frac{\sin(k r_{12})}{k r_{12}} + \biggr(\hat{ \boldsymbol{\mu}}_1 \cdot \hat{ \boldsymbol{\mu}}_2 - 3(\hat{ \boldsymbol{\mu}}_1 \cdot \hat{\boldsymbol{r}}_{12})(\hat{ \boldsymbol{\mu}}_2 \cdot \hat{\boldsymbol{r}}_{12}) \biggr)\biggr(\frac{\cos(k r_{12})}{(k r_{12})^2} - \frac{\sin(k r_{12})}{(k r_{12})^3}\biggr) \biggr],
    \end{align}
\end{subequations}
where $\alpha$ is the combined Debye-Waller/Franck-Condon factor, $\gamma_0$ is the total decay rate from the electronic excited state $\ket{e_j}$, $\hat{\boldsymbol{r}}_{12}=(\boldsymbol{r}_{2}-\boldsymbol{r}_{1})/r_{12}$ is the unit vector pointing from the position $\boldsymbol{r}_{1}$ of emitter $j=1$ to the position $\boldsymbol{r}_{2}$ of emitter $j=2$, $r_{12}=|\boldsymbol{r}_{2}-\boldsymbol{r}_{1}|$ is the distance between the emitters, and $k=n (\omega_1 + \omega_2 )/(2 c)$ is the wavenumber of the arithmetic mean of the transition frequency of the two emitters, with $c$ the speed of light in vacuum and $n$ the refractive index of the host matrix. Additionally, $\hat{\boldsymbol{\mu}}_j = \boldsymbol{\mu}_j / |\boldsymbol{\mu}_j|$ is the unit vector of the transition dipole moment of emitter $j$ (corresponding to the transition between the pure electronic excited state $\ket{e_j}$ and the pure electronic ground state $\ket{g_j}$). The norm of $\boldsymbol{\mu}_j$ is $|\boldsymbol{\mu}_j | = \sqrt{\alpha \gamma_0 3\pi \varepsilon_0 \hbar n^2 /k^3}$, with $\varepsilon_0$ the vacuum permittivity. The expressions in Eq. (\ref{EqSM:complete_expressions_coupling}) include retardation effects \cite{Stokes_NJP_2018}, as well as the combined Debye-Waller/Franck-Condon factor $\alpha$ \cite{Trebbia_NatComms_2022, JuanDelgado}. The latter corresponds to the proportion of photons emitted into the Zero-Phonon Line (ZPL) over the total number of photons emitted from the quantum emitter. Thus, this factor accounts in an effective manner for the phonons of the host matrix in which the emitters are embedded, as well as for the internal vibrational modes of the quantum emitters that are not taken explicitly into account. 

\begin{figure}[t] 
	\begin{center}
		\includegraphics[width=0.95\textwidth]{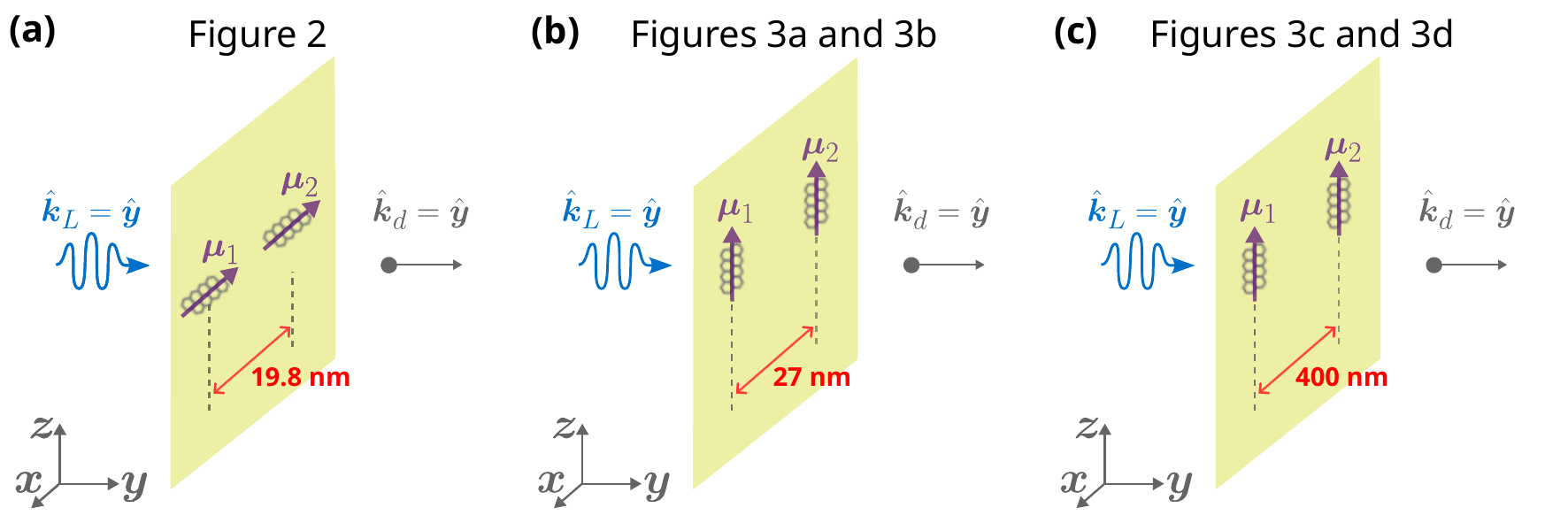}%
		\caption{Schematic representations of the molecular pairs used in the simulations of the intensity correlation presented in the main text. (a) J-aggregate molecular configuration simulated in Fig. 2 of the main text, where the transition dipole moments $\boldsymbol{\mu}_j$ are oriented in the $x$-direction and the molecules are separated by $19.8$ nm in the x-direction. (b) Molecular pair simulated in Figs. 3a and 3b of the main text, with the transition dipole moments oriented along the $z$-direction and forming a molecular H-aggregate with intermolecular distance of $27$ nm in the $x$-direction. (c) Molecular pair with large separation distance in the $x$-direction ($400$ nm) and in a H-aggregate configuration (with the transition dipole moments aligned along the $z$-direction), which is used in the simulations in Figs. 3c and 3d of the main text. In all panels, we consider incidence and detection of light in the normal direction to the $xz$-plane where the transition dipole moments are contained.}  
        \label{FigureSM:configurations}  
	\end{center}
\end{figure}

\section{SII. Parameters used in simulations}
\begin{table}[b]
\centering
    \begin{tabular}{|c|c|c|c|c|c|c|c|c|c|c|c|c|c|}
        \hline
         Figure & $\boldsymbol{\mu}_j/|\boldsymbol{\mu}_j|$ & $\boldsymbol{r}_1$ & $\boldsymbol{r}_2$ [nm] & $\gamma_0 / (2\pi)$ [MHz] & $\alpha$ & $V/\gamma_0$ & $\gamma_{ij}/\gamma_0$ & $\delta/\gamma_0$ & $\Omega_j/\gamma_0$ & $\boldsymbol{k}_L/|\boldsymbol{k}_L|$ & $\phi_{\text{ZPL},\text{St}}$ & $\hbar\omega_v$ [meV] & $1/\gamma_v$ [ps]\\
         \hline
         Fig. 2      & $\hat{\boldsymbol{x}}$ & $0$ & $19.8 \boldsymbol{\hat{x}}$ & $21.5$ & $0.3$ & $-17$ & $0.3$ & $14$ & $3.15$ & $\hat{\boldsymbol{y}}$ & $0$ & $31.86$ & $10$ \\
         \hline
         Fig. 3a,b & $\hat{\boldsymbol{z}}$ & $0$ & $27 \boldsymbol{\hat{x}}$ & $21.5$ & $0.3$ & $2.98$ & $0.29$ & $5$ & $1.5$ & $\hat{\boldsymbol{y}}$ & $0$ &$31.86$ & $10$ \\
         \hline
         Fig. 3c,d & $\hat{\boldsymbol{z}}$ & $0$ & $400 \boldsymbol{\hat{x}}$ & $21.5$ & $0.3$ & $-0.04$ & $-10^{-3}$ & $5$ & $1.5$ & $\hat{\boldsymbol{y}}$ & $0$ &$31.86$ & $10$ \\
        \hline
    \end{tabular}
    \caption{Molecular and laser parameters used in the simulations in Fig. 2 (second row), in Figs. 3a and 3b (third row) and in Figs. 3c and 3d (fourth row) in the main text.}
    \label{Table}
\end{table}
In this section we provide all the parameters of the different molecular configurations simulated in the main text (see Table \ref{Table}), as well as schematic representations of each of these configurations. In all cases we fix $\gamma_0 /(2\pi) = 21.5$ MHz, $\alpha=0.3$, $n=1.5$, and $\omega_0 = (\omega_1 + \omega_2)/2$ corresponding to a wavelength of 618 nm \cite{Trebbia_NatComms_2022}. Additionally, we consider normal incidence and detection of light with respect to the $xz$-plane where the transition dipole moments $\boldsymbol{\mu}_j$ are contained, as sketched in Fig. \ref{FigureSM:configurations}, with both the laser wave vector $\boldsymbol{k}_L$ and the detection direction $\boldsymbol{k}_d$ oriented along the $y$-axis. The molecules are excited by the component of the incident electric field that is parallel to the direction of the transition dipole moments. In Fig. \ref{FigureSM:configurations}a we plot the molecular configuration corresponding to the simulations in Fig. 2 of the main text, where the transition dipole moments are oriented along the $x$-direction and we fix $\boldsymbol{r}_{12}=\boldsymbol{\hat{x}} 19.8$ nm (thus, in a J-aggregate configuration, which is characterized by transition dipole moments aligned in the same direction than $\boldsymbol{r}_{12}$), which yields $V=-17\gamma_0$ and $\gamma_{12}=0.3\gamma_0$. Next, Fig. \ref{FigureSM:configurations}b shows a schematic of the molecular pair simulated in Figs. 3a and 3b of the main text, with the transition dipole moments oriented along the $z$-direction and $\boldsymbol{r}_{12}=\boldsymbol{\hat{x}} 27$ nm (H-aggregate configuration, with transition dipole moments aligned in the direction perpendicular to $\boldsymbol{r}_{12}$), leading to $V=2.98\gamma_0$ and $\gamma_{12}=0.29\gamma_0$. 
Last, we represent in Fig. \ref{FigureSM:configurations}c the molecular pair simulated in Figs. 3c and 3d of the main text, where the molecules are considered to be separated by $400$ nm and in a H-aggregate configuration, which results in $V=-0.04\gamma_0$ and $\gamma_{12}=-10^{-3}\gamma_0$. The values of $V$, $\gamma_{12}$ and $\delta$ in Figs. 2 and 3a,b are obtained by fitting the experimental excitation spectra at different laser intensities, see Section SVI. Additional measurements of the excitation spectra are performed at different polarizer angles \cite{Trebbia_NatComms_2022}, which yields that the dipoles are parallel. $\boldsymbol{r}_1$ and $\boldsymbol{r}_2$ are then extracted from the values of $V$ and $\gamma_{12}$ and taking into account the parallel orientation of the dipoles.

Moreover, the value of Rabi frequency used in the simulations shown in Figs. 2a-c of the main text is $\Omega_j = 3.15 \gamma_0$, which can be estimated from the experimental laser intensity used in these experiment ($I=100$ $W/cm^2$) and the corresponding saturation intensity of each molecule ($I_{sat}=5.1$ $W/cm^2$). These values of experimental intensities correspond to a laser linearly polarized in the same direction as the transition dipole moments, and thus they are related to the Rabi frequency through the expression $\Omega_j=\gamma_0 \sqrt{I/(2I_{sat})}$ \cite{Steck_book_2007}. On the other hand, in the simulations in Figs. 3a and 3b of the main text we use $\Omega_j = 1.5 \gamma_0$, estimated from the experimental values $I=60$ $W/cm^2$ and $I_{sat}=14$ $W/cm^2$.


\section{SIII. Typical framework with two-level systems}
In this section we review the typical framework describing the emitters as two-level systems (TLSs) and we show that it results in almost identical intensity  correlation of ZPL photons as the model that we propose in the main text. However, we emphasize that the TLS framework introduced here cannot address the correlation of Stokes-shifted photons, in contrast to the model that we have presented in the main text including the vibrational levels and which allows for addressing the correlation of Stokes-shifted photons and also of ZPL photons.

Within the TLS framework, the Hamiltonian can be written as 
\begin{equation}
    \tH = \hbar \sum_{j=1}^2 \biggr[ \frac{\Delta_j}{2} ( \ket{e_j}\bra{e_j}-\ket{g_j}\bra{g_j}) \biggr] + \H_I + \H_P ,
\end{equation}
where the interaction Hamiltonian is $\H_{I}=\hbar V(\sd_1 \s_2 + \s_1 \sd_2)$ and the pumping Hamiltonian is $\H_{P}=-\frac{\hbar}{2} \sum_{j=1}^2 (\Omega_j \sd_j + \Omega_j^* \s_j)$, with $\s_j = \ket{g_j}\bra{e_j}$ and $\sd_j = \ket{e_j}\bra{g_j}$. Here, we use the tilde symbol in the Hamiltonian $\tH$ to distinguish this description from the model including vibrations that we use in the main text. In this case, the Markovian master equation governing the dynamics of the density matrix $\tdm$ takes the form
\begin{equation} \label{EqSM:master_equation_TLS}
    \frac{d}{dt}\tdm =-\frac{i}{\hbar}[\tH, \tdm] + \sum_{j=1}^2 \biggr(\frac{\gamma_{0}}{2}\mathcal{D}[\s_j] +  \sum_{k\neq j} \frac{\gamma_{jk}}{2} \mathcal{D}[\s_j , \s_k]  \biggr)\tdm .
\end{equation}

The correlation of ZPL photons can be obtained from Eq. (3) of the main text using the electric field operator
\begin{equation} \label{EqSM:electric_field_TLS_framework}
    \E_{ZPL}^{(+)} (t) =  \E_{ZPL,1}^{(+)} (t) + \E_{ZPL,2}^{(+)} (t)  = \xi_{ZPL} \biggr( \hat{\sigma}_1 (t) + e^{i\phi_{ZPL}} \hat{\sigma}_2 (t) \biggr) , 
\end{equation}
where $\E_{ZPL,1}^{(+)} (t) = \xi_{ZPL} \hat{\sigma}_1 (t)$ and $\E_{ZPL,2}^{(+)} (t) = \xi_{ZPL} \hat{\sigma}_2 (t) e^{i\phi_{ZPL}}$. We have verified that the maximum difference between the correlation of ZPL photons obtained with the model in the main text and the TLS framework presented in this appendix is $|\tilde{g}_{ZPL}^{(2)}(\tau)-{g}_{ZPL}^{(2)}(\tau)|/{g}_{ZPL}^{(2)}(\tau) \approx 7\cdot 10^{-4}$. This difference occurs in the configuration in Fig. 2c of the main text and at delay $\tau=0$.

\section{SIV. Conditional-probability approach to the correlation of Stokes-shifted photons}
\begin{figure}[t] 
	\begin{center}
		\includegraphics[width=0.65\textwidth]{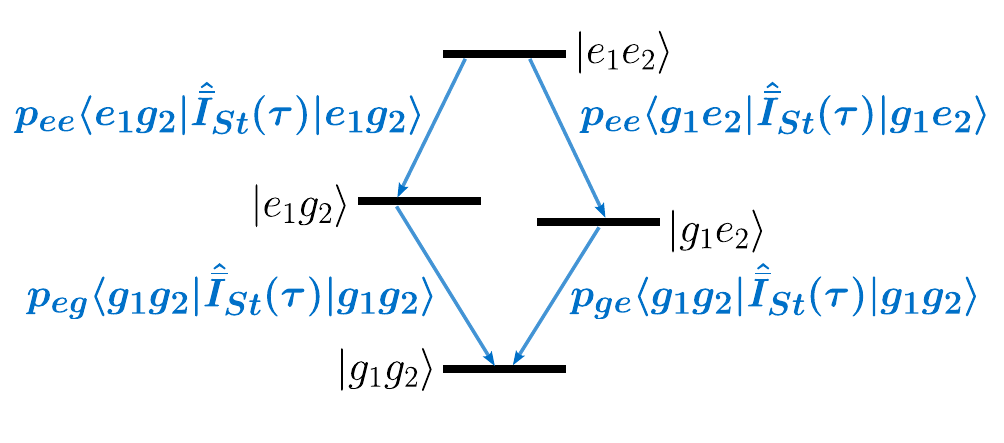}%
		\caption{Schematic representation of the contributions to the numerator of the Stokes-shifted correlation correlation $\bar{g}_{\text{St}}^{(2)} (\tau)$ within the conditional-probability approach [Eq. (\ref{EqSM:g2_statistical_approach})]. Each contribution is given by the product of a steady-state population and a diagonal element of the operator $\hat{\bar{I}}_{\text{St}}(\tau)$, which describes the Stokes-shifted intensity within this approach. Specifically, the steady-state population corresponds to the initial state of the one-photon transition, whereas the diagonal element of the operator $\hat{\bar{I}}_{\text{St}}(\tau)$ is given by the final state of the one-photon transition.}
        \label{FigureSM:statistical_approach}  
	\end{center}
\end{figure}
The Stokes-shifted correlation $g_{\text{St,d}}^{(2)}(\tau)$ in the main text was obtained by neglecting the role of quantum coherence within the new model that we propose. $g_{\text{St,d}}^{(2)}(\tau)$ gives almost 
identical results to the approach adopted in Refs. \cite{Hettich_Science_2002, Trebbia_NatComms_2022} to describe the statistics of Stokes-shifted photons, which we call \textit{conditional-probability approach}. This approach is not used in our work, but we describe it here for reference. 

Within the conditional-probability approach, the calculation of the intensity correlation $g^{(2)}(\tau)$ is based on the reset matrix formalism \cite{Hegerfeldt_PRA_1993,Beige_PRA_1998}. More specifically, $g^{(2)}(\tau)$ is calculated from the conditional probability of emitting a photon at time $t+\tau$, assuming that the system has emitted a photon at an earlier time $t$. At the steady state, the reset matrix formalism yields
\begin{equation} \label{EqSM:reset_matrix_g2}
g^{(2)}(\tau)=\frac{\text{Tr}\{\hat{I}(\tau)\dm_{\text{reset}}\}}{\braket{\hat{I}(0)}_{ss}} .
\end{equation}
Here, the denominator is the steady-state expected value  of the light intensity operator $\hat{I}$, whereas the numerator represents the expected value of the Heisenberg-picture operator $\hat{I}(\tau)$ for an initial state $\dm_{\text{reset}}$. Importantly, $\dm_{\text{reset}}$ is referred to as reset density matrix and corresponds to the density matrix of the system right after emitting a photon, provided that the system was in the steady state before emitting the photon (see below for details) .

Furthermore, within the conditional-probability approach, the correlation of the Stokes-Shifted photons from two interacting emitters have been described using the two-level-system framework discussed in Section SIII of this Supplemental Material \cite{Hettich_Science_2002, Hettich_thesis_2002, Trebbia_NatComms_2022}. By using Eq. (\ref{EqSM:electric_field_TLS_framework}), we obtain that the intensity operator is given by
\begin{equation} \label{EqSM:intensity_operator_ZPL_statistical_approach}
\begin{split}
    \hat{I} (\tau) &= \biggr[\hat{E}_{ZPL,1}^{(-)} (\tau) + \hat{E}_{ZPL,2}^{(-)} (\tau)\biggr] \biggr[\hat{E}_{ZPL,1}^{(+)} (\tau) + \hat{E}_{ZPL,2}^{(+)} (\tau)\biggr] = |\xi_{ZPL}|^2   \biggr[\sd_1 (\tau) + \sd_2 (\tau)\biggr] \biggr[\s_1 (t) + \s_2 (t)\biggr] \\
    &=|\xi_{ZPL}|^2 \biggr[ 2\ket{e_1 e_2}\bra{e_1 e_2} (\tau) + \ket{e_1 g_2}\bra{e_1 g_2} (\tau) + \ket{g_1 e_2}\bra{g_1 e_2} (\tau) + \ket{e_1 g_2}\bra{g_1 e_2} (\tau) + \ket{g_1 e_2}\bra{e_1 g_2} (\tau) \biggr].
\end{split}
\end{equation}
At this point, this intensity operator describes the intensity of light emitted in the ZPL (and not yet the Stokes-shifted emission) and, notably, it accounts for the first-order coherence between the light radiated from the two emitters through the terms that mix the electric field operators of both emitters \cite{Glauber_PR_1963}, which are $\hat{E}_{ZPL,1}^{(-)} (\tau)\hat{E}_{ZPL,2}^{(+)} (\tau)$ and $\hat{E}_{ZPL,2}^{(-)} (\tau)\hat{E}_{ZPL,1}^{(+)} (\tau)$. These two terms lead to the $ \ket{e_1 g_2}\bra{g_1 e_2} (\tau)$ and $\ket{g_1 e_2}\bra{e_1 g_2} (\tau) $ terms in the second line of Eq. (\ref{EqSM:intensity_operator_ZPL_statistical_approach}), which are off-diagonal in the basis $\{\ket{e_1 e_2}, \ket{e_1 g_2}, \ket{g_1 e_2}, \ket{g_1 g_2}\}$. In contrast, the other terms in the second line of Eq. (\ref{EqSM:intensity_operator_ZPL_statistical_approach}) come from the intensity of the light emitted independently from each emitter [i.e., from $\hat{E}_{ZPL,1}^{(-)} (\tau)\hat{E}_{ZPL,1}^{(+)} (\tau)$ and $\hat{E}_{ZPL,2}^{(-)} (\tau)\hat{E}_{ZPL,2}^{(+)} (\tau)$] and are diagonal in this same basis $\{\ket{e_1 e_2}, \ket{e_1 g_2}, \ket{g_1 e_2}, \ket{g_1 g_2}\}$. To address the Stokes-shifted emission instead of the ZPL emission, Refs. \cite{Hettich_Science_2002, Hettich_thesis_2002, Trebbia_NatComms_2022} assume that quantum coherence does not play any role in the emission of Stokes-shifted photons and identify the Stokes-shifted intensity operator with the diagonal elements terms in the second line of Eq. (\ref{EqSM:intensity_operator_ZPL_statistical_approach}). Thus,
\begin{equation} \label{EqSM:intensity_operator_Stokes_statistical_approach}
    \hat{\bar{I}}_{\text{St}} (\tau) \propto 2\ket{e_1 e_2}\bra{e_1 e_2} (\tau) + \ket{e_1 g_2}\bra{e_1 g_2} (\tau) + \ket{g_1 e_2}\bra{g_1 e_2} (\tau) ,
\end{equation}
where we use the bar symbol to stress that this is the expression used in the conditional-probability approach. In the steady state, the expected value of the operator in Eq. (\ref{EqSM:intensity_operator_Stokes_statistical_approach}) thus becomes
\begin{equation} \label{EqSM:Stokes-shifted_I_stateistica_approach}
    \braket{\hat{\bar{I}}_{\text{St}}(0)}_{ss} \propto 2p_{ee} + p_{eg} + p_{ge},
\end{equation}
with $p_{ee}$ the steady-state population of the doubly excited state $\ket{e_1 e_2}$, $p_{eg}$ the steady-state population of $\ket{e_1 g_2}$, and $p_{ge}$ the steady-state population of $\ket{g_1 e_2}$. 

Moreover, within the conditional-probability approach, the off-diagonal elements of the steady-state density matrix in the $\{\ket{e_1 e_2}, \ket{e_1 g_2}, \ket{g_1 e_2}, \ket{g_1 g_2}\}$ basis are also neglected. These elements are identified with the quantum coherence of the state of the system \cite{Hettich_thesis_2002}. Thus, the steady-state density matrix results in
\begin{equation}
    \hat{\bar{ \rho}}_{ss} = p_{ee} \ket{e_1 e_2}\bra{e_1 e_2} + p_{eg} \ket{e_1 g_2}\bra{e_1 g_2} + p_{ge} \ket{g_1 e_2}\bra{g_1 e_2} + p_{gg} \ket{g_1 g_2}\bra{g_1 g_2}  .
\end{equation}
The resulting reset density matrix becomes \cite{Hettich_thesis_2002}
\begin{equation} \label{EqSM:reset_matrix}
    \dm_{\text{reset}} = \frac{p_{ee} \ket{e_1 g_2}\bra{e_1 g_2} + p_{ee} \ket{g_1 e_2}\bra{g_1 e_2} + p_{eg} \ket{g_1 g_2}\bra{g_1 g_2} + p_{ge} \ket{g_1 g_2}\bra{g_1 g_2}}{2p_{ee} + p_{eg} + p_{ge}} ,
\end{equation}
where the four terms in the numerator can be interpreted from the one-photon transitions between states of the uncoupled basis $\{\ket{e_1 e_2}, \ket{e_1 g_2}, \ket{g_1 e_2}, \ket{g_1 g_2}\}$, which are schematically represented in Fig. \ref{FigureSM:statistical_approach}. Specifically, the doubly-excited state $\ket{e_1 e_2}$ can decay either to $\ket{e_1 g_2}$ or to $\ket{e_1 g_2}$, which would reset the state to $p_{ee}\ket{e_1 g_2}\bra{e_1 g_2}$ or to $p_{ee}\ket{g_1 e_2}\bra{g_1 e_2}$. Similarly, both $\ket{e_1 g_2}$ and $\ket{g_1 e_2}$ can decay to the ground state $\ket{g_1 g_2}$, reseting the state to $p_{eg} \ket{g_1 g_2}\bra{g_1 g_2}$ and $p_{ge} \ket{g_1 g_2}\bra{g_1 g_2}$, respectively. We emphasize that the conditional-probability approach considers the uncoupled basis $\{\ket{e_1 e_2}, \ket{e_1 g_2}, \ket{g_1 e_2}, \ket{g_1 g_2}\}$ because the emitters are assumed to couple only through the ZPL and, consequently, the emission of Stokes-shifted photons projects the system into a localized state of this basis \cite{Hettich_thesis_2002}. 

Finally, by substituting Eqs. (\ref{EqSM:Stokes-shifted_I_stateistica_approach}) and (\ref{EqSM:reset_matrix}) into Eq. (\ref{EqSM:reset_matrix_g2}), we obtain that the Stokes-shifted photon correlation within this conditional-probability approach is given by
\begin{equation} \label{EqSM:g2_statistical_approach}
    \bar{g}_{\text{St}}^{(2)} (\tau) = \frac{p_{ee} \bra{e_1 g_2}\hat{I}_{\text{St}}(\tau)\ket{e_1 g_2}+p_{ee} \bra{g_1 e_2}\hat{I}_{\text{St}}(\tau)\ket{g_1 e_2} + p_{eg} \bra{g_1 g_2}\hat{I}_{\text{St}}(\tau)\ket{g_1 g_2} + p_{ge} \bra{g_1 g_2}\hat{I}_{\text{St}}(\tau)\ket{g_1 g_2}}{\braket{\hat{I}_{\text{St}}(0)}_{ss} [2p_{ee} + p_{eg} + p_{ge}]} .
\end{equation}
This expression has an intuitive statistical interpretation, based on the contribution of each allowed one-photon transition (in the basis $\{\ket{e_1 e_2}, \ket{e_1 g_2}, \ket{g_1 e_2}, \ket{g_1 g_2}\}$) to increase the probability of having coincidences in the Hanbury-Brown Twiss interferometer used to measure the intensity correlation. More specifically, the numerator in Eq. (\ref{EqSM:g2_statistical_approach}) is decomposed in four terms, given by products of steady-state populations and elements of the intensity operator. The steady-state population corresponds to the initial state of the one-photon transition, whereas the diagonal element of the intensity operator corresponds to the final state of the transition, as schematically represented in Fig. \ref{FigureSM:statistical_approach}. For example, $p_{ee} \bra{e_1 g_2}\hat{\bar{I}}_{\text{St}}(\tau)\ket{e_1 g_2}$ corresponds to the contribution to the intensity correlation of the one-photon transition from $\ket{e_1 e_2}$ to $\ket{e_1 g_2}$. We have verified that Eq. (\ref{EqSM:g2_statistical_approach}) yields almost identical results as $g_{\text{St,d}}^{(2)}(\tau)$, which is obtained by neglecting the role of coherence within the model that we propose in this Letter. More specifically, the difference between them is in general $< 1\%$, with the largest difference $\approx 3 \%$ and obtained for the configuration in Fig. 2b. Indeed, we show in Section SV of this Supplemental Material that the analytical expression of $g_{\text{St,d}}^{(2)}(\tau)$ can be approximated by the expression of $\bar{g}_{\text{St}}^{(2)} (\tau)$ in Eq. (\ref{EqSM:g2_statistical_approach}).

\section{SV. Analytical expression of the correlation of Stokes-shifted photons}
We detail in this section the derivation of the analytical expression of the intensity correlation of Stokes-shifted photons presented in Eq. (4) of the main text. First, we substitute the electric field operator $\E_{\text{St}}^{(+)} (t) / \xi_{\text{St}} = \ket{v_1}\bra{e_1} (t) + \ket{v_2}\bra{e_2} (t)$ into the general expression of the intensity (or second-order) correlation in Eq. (3) of the main text, which yields
\begin{equation} \label{Eq:correlation_Stokes_first_analytical}
    g_{\text{St}}^{(2)}(\tau)= \frac{G_{\text{St}}^{(2)}(\tau)}{\braket{\hat{I}_{\text{St}}(0)}^2} =  \frac{|\xi_{\text{St}}|^2 \braket{(\ket{e_1}\bra{v_1} + \ket{e_2}\bra{v_2}) I_{\text{St}} (\tau)(\ket{v_1}\bra{e_1} + \ket{v_2}\bra{e_2})}}{\braket{\hat{I}_{\text{St}}(0)}^2} .
\end{equation}
Here, we have introduced the operator
\begin{equation} \label{EqSM:I_stokes}
    \hat{I}_{\text{St}} (t)=\E_{\text{St}}^{(-)} (t)\E_{\text{St}}^{(+)} (t)= |\xi_{\text{St}}|^2 \biggr(\ket{e_1}\bra{v_1} (t) + \ket{e_2}\bra{v_2} (t)\biggr)\biggr(\ket{v_1}\bra{e_1} (t) + \ket{v_2}\bra{e_2} (t)\biggr) ,
\end{equation}
whose steady-state expected value $\braket{\hat{I}_{\text{St}}(0)}$ is proportional to light intensity and is given by
\begin{equation} \label{EqSM:mean_value_intensity_Stokes}
    \braket{\hat{I}_{\text{St}}(0)} = |\xi_{\text{St}}|^2 \biggr( 2 p_{ee} + p_{eg} + p_{ge} + p_{ev} +p_{ve} + 2\text{Re}\rho_{ev,ve} \biggr) .
\end{equation}
where $p_{ab}=\braket{a_1 b_2|\dm_{ss}| a_1 b_2}$ is the steady-state population of the state $\ket{a_1 b_2}$ and $\rho_{ab,cd}=\braket{a_1 b_2|\dm_{ss} | c_1 d_2}$ is an off-diagonal element of the steady-state density matrix, with $a,b,c,d\in\{e,g,v\}$.

Next, recalling that the steady-state value of any operator $\hat{A}$ is given by $\braket{\hat{A}}_{ss}=\text{Tr}(\hat{A} \dm_{ss})$ and using the cyclic property of trace, we rewrite the numerator in Eq. (\ref{Eq:correlation_Stokes_first_analytical}) as
\begin{equation} \label{EqSM:complete_G2_Stokes}
\begin{split}
    G_{\text{St}}^{(2)}(\tau) / |\xi_{\text{St}}|^4 &= p_{ee}\bra{v_1 e_2}\hat{I}_{\text{St}}(\tau)\ket{v_1 e_2} + p_{ee} \bra{e_1 v_2}\hat{I}_{\text{St}}(\tau)\ket{e_1 v_2} + p_{eg} \bra{v_1 g_2}\hat{I}_{\text{St}}(\tau)\ket{v_1 g_2} \\
    &+ p_{ge} \bra{g_1 v_2}\hat{I}_{\text{St}}(\tau)\ket{g_1 v_2} + p_{ev} \bra{v_1 v_2} \hat{I}_{\text{St}}(\tau) \ket{v_1 v_2} + p_{ve}\bra{v_1 v_2} \hat{I}_{\text{St}}(\tau) \ket{v_1 v_2} \\
    &+ (\bra{v_1 e_2}\hat{I}_{\text{St}}(\tau)\ket{e_1 v_2} + \bra{e_1 v_2}\hat{I}_{\text{St}}(\tau)\ket{v_1 e_2}) p_{ee} \\
    &+\rho_{eg,ee} [\bra{v_1 e_2 }\hat{I}_{\text{St}}(\tau)\ket{v_1 g_2} + \bra{e_1 v_2 }\hat{I}_{\text{St}}(\tau)\ket{v_1 g_2}] + \rho_{ge,ee} [\bra{v_1 e_2 }\hat{I}_{\text{St}}(\tau)\ket{g_1 v_2} + \bra{e_1 v_2 }\hat{I}_{\text{St}}(\tau)\ket{g_1 v_2}] \\
    &+ \rho_{ee,eg} [\bra{v_1 g_2 }\hat{I}_{\text{St}}(\tau)\ket{e_1 v_2} + \bra{v_1 g_2 }\hat{I}_{\text{St}}(\tau)\ket{v_1 e_2}] + \rho_{ee,ge} [\bra{g_1 v_2 }\hat{I}_{\text{St}}(\tau)\ket{e_1 v_2} + \bra{g_1 v_2 }\hat{I}_{\text{St}}(\tau)\ket{v_1 e_2}] \\
        &+ (\rho_{ee,ve} + \rho_{ee,ev})[\bra{v_1 v_2 }\hat{I}_{\text{St}}(\tau)\ket{v_1 e_2} + \bra{v_1 v_2 }\hat{I}_{\text{St}}(\tau)\ket{e_1 v_2}] \\
        &+ (\rho_{ve,ee} + \rho_{ev,ee})[\bra{e_1 v_2 }\hat{I}_{\text{St}}(\tau)\ket{v_1 v_2} + \bra{v_1 e_2 }\hat{I}_{\text{St}}(\tau)\ket{v_1 v_2}] \\
        &+ (\rho_{eg,ev} + \rho_{eg,ve}) \bra{v_1 v_2} \hat{I}_{\text{St}}(\tau) \ket{v_1 g_2} +  (\rho_{ge,ev} + \rho_{ge,ve}) \bra{v_1 v_2} \hat{I}_{\text{St}}(\tau) \ket{g_1 v_2} \\
        &+ (\rho_{ev,eg} + \rho_{ve,eg}) \bra{v_1 g_2} \hat{I}_{\text{St}}(\tau) \ket{v_1 v_2} +  (\rho_{ev,ge} + \rho_{ve,ge}) \bra{g_1 v_2} \hat{I}_{\text{St}}(\tau) \ket{v_1 v_2} \\
        &+ \rho_{eg,ge} \bra{g_1 v_2} \hat{I}_{\text{St}}(\tau) \ket{v_1 g_2} + \rho_{ge,eg} \bra{v_1 g_2} \hat{I}_{\text{St}}(\tau) \ket{g_1 v_2} + (\rho_{ev,ve } + \rho_{ve,ev})\bra{v_1 v_2} \hat{I}_{\text{St}}(\tau) \ket{v_1 v_2}.
\end{split}    
\end{equation}
\begin{figure}[!t] 
	\begin{center}
		\includegraphics[width=0.85\textwidth]{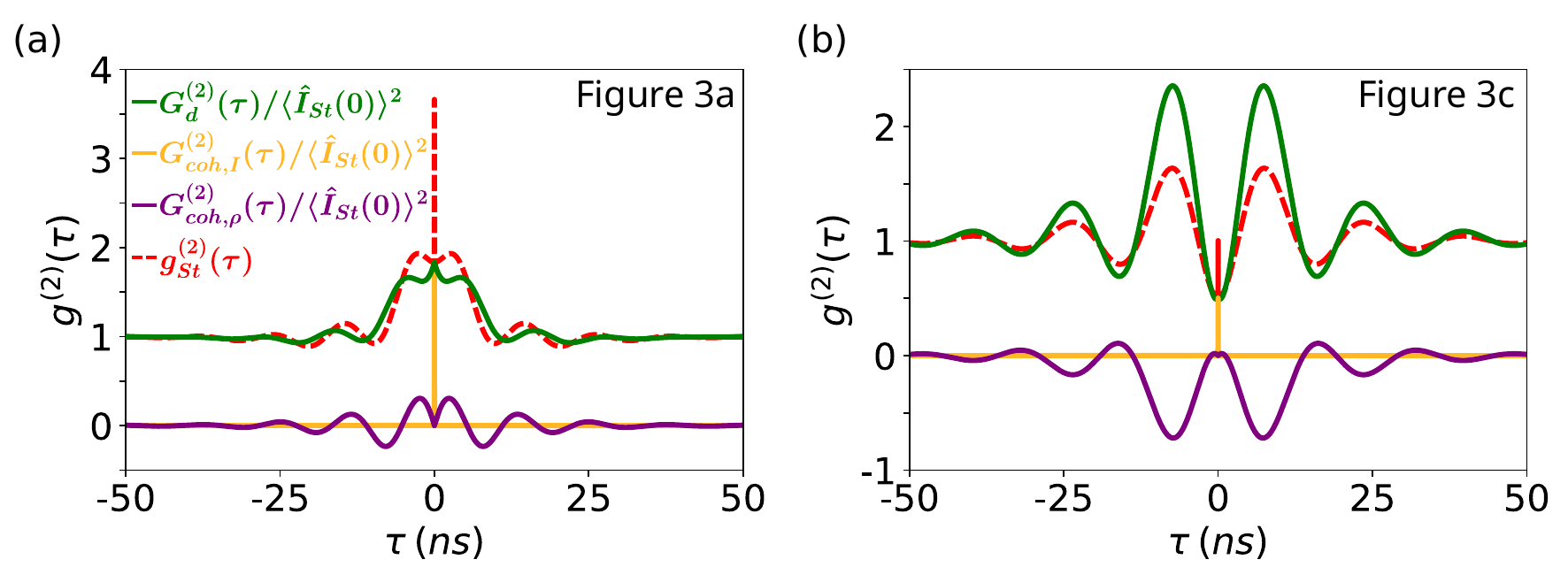}%
		\caption{Dependence on the time delay $\tau$ of the different terms in which $G_{\text{St}}^{(2)}(\tau)$ is decomposed in Eq. (\ref{EqSM:G2_st_decomposition}). Solid green lines correspond to $G_{d}^{(2)}(\tau)/\braket{\hat{I}_{\text{St}}(0)}^2$, solid orange lines to $G_{coh,I}^{(2)}(\tau)/\braket{\hat{I}_{\text{St}}(0)}^2$, solid purple lines to $G_{coh,\rho}^{(2)}(\tau)/\braket{\hat{I}_{\text{St}}(0)}^2$, and dashed red lines to the total correlation of Stokes-shifted photons $g_{\text{St}}^{(2)}(\tau)$ (given by the sum of the three former quantities). We simulate in (a) the molecular configuration used in Fig. 3a of the main text (all the parameters are specified in the third row of Table \ref{Table}), and in (b) the molecular configuration used in Fig. 3c of the main text (all the parameters are specified in the fourth row of Table \ref{Table}).   }
        \label{FigureSM:G2_decomposition}  
	\end{center}
\end{figure}
This expression can be decomposed into three different contributions 
\begin{equation} \label{EqSM:G2_st_decomposition}
    G_{\text{St}}^{(2)}(\tau) 
 = G_{d}^{(2)}(\tau) + G_{coh,I}^{(2)}(\tau) + G_{coh,\rho}^{(2)}(\tau) .
\end{equation}
The first contribution is given by the six terms in the first two lines of Eq. (\ref{EqSM:complete_G2_Stokes}) ,
\begin{equation} \label{EqSM:Gd}
\begin{split}
    G_{d}^{(2)}(\tau) / |\xi_{\text{St}}|^4 &= p_{ee}\bra{v_1 e_2}\hat{I}_{\text{St}}(\tau)\ket{v_1 e_2} + p_{ee} \bra{e_1 v_2}\hat{I}_{\text{St}}(\tau)\ket{e_1 v_2} + p_{eg} \bra{v_1 g_2}\hat{I}_{\text{St}}(\tau)\ket{v_1 g_2} \\
    &+ p_{ge} \bra{g_1 v_2}\hat{I}_{\text{St}}(\tau)\ket{g_1 v_2} + p_{ev} \bra{v_1 v_2} \hat{I}_{\text{St}}(\tau) \ket{v_1 v_2} + p_{ve}\bra{v_1 v_2} \hat{I}_{\text{St}}(\tau) \ket{v_1 v_2}  ,
\end{split}
\end{equation}
involving only diagonal elements of $\dm_{ss}$ and of $\hat{I}_{\text{St}}(\tau)$. The second contribution includes the terms in the third line of Eq. (\ref{EqSM:complete_G2_Stokes}) 
\begin{equation}
     G_{coh,I}^{(2)}(\tau)/ |\xi_{\text{St}}|^4  = p_{ee} (\bra{v_1 e_2}\hat{I}_{\text{St}}(\tau)\ket{e_1 v_2} + \bra{e_1 v_2}\hat{I}_{\text{St}}(\tau)\ket{v_1 e_2})  ,
\end{equation}
which involves diagonal elements of the density matrix (populations) and off-diagonal elements of the operator $\hat{I}_{\text{St}}(\tau)$. Thus, $G_{coh,I}^{(2)}(\tau)$ is affected by the first-order coherence of the electric field emitted from the two emitters, see Section SIV of this Supplemental Material. The rest of terms in Eq. (\ref{EqSM:complete_G2_Stokes}) include off-diagonal elements of the steady-state density matrix, as well as of the intensity operator. In this way, we define 
\begin{equation} \label{EqSM:G2_coh_rho}
    \begin{split}
        G_{coh,\rho}^{(2)}(\tau) / |\xi_{\text{St}}|^4 &= \rho_{eg,ee} [\bra{v_1 e_2 }\hat{I}_{\text{St}}(\tau)\ket{v_1 g_2} + \bra{e_1 v_2 }\hat{I}_{\text{St}}(\tau)\ket{v_1 g_2}] + \rho_{ge,ee} [\bra{v_1 e_2 }\hat{I}_{\text{St}}(\tau)\ket{g_1 v_2} + \bra{e_1 v_2 }\hat{I}_{\text{St}}(\tau)\ket{g_1 v_2}] \\
        &+ \rho_{ee,eg} [\bra{v_1 g_2 }\hat{I}_{\text{St}}(\tau)\ket{e_1 v_2} + \bra{v_1 g_2 }\hat{I}_{\text{St}}(\tau)\ket{v_1 e_2}] + \rho_{ee,ge} [\bra{g_1 v_2 }\hat{I}_{\text{St}}(\tau)\ket{e_1 v_2} + \bra{g_1 v_2 }\hat{I}_{\text{St}}(\tau)\ket{v_1 e_2}] \\
        &+ (\rho_{ee,ve} + \rho_{ee,ev})[\bra{v_1 v_2 }\hat{I}_{\text{St}}(\tau)\ket{v_1 e_2} + \bra{v_1 v_2 }\hat{I}_{\text{St}}(\tau)\ket{e_1 v_2}] \\
        &+ (\rho_{ve,ee} + \rho_{ev,ee})[\bra{e_1 v_2 }\hat{I}_{\text{St}}(\tau)\ket{v_1 v_2} + \bra{v_1 e_2 }\hat{I}_{\text{St}}(\tau)\ket{v_1 v_2}] \\
        &+ (\rho_{eg,ev} + \rho_{eg,ve}) \bra{v_1 v_2} \hat{I}_{\text{St}}(\tau) \ket{v_1 g_2} +  (\rho_{ge,ev} + \rho_{ge,ve}) \bra{v_1 v_2} \hat{I}_{\text{St}}(\tau) \ket{g_1 v_2} \\
        &+ (\rho_{ev,eg} + \rho_{ve,eg}) \bra{v_1 g_2} \hat{I}_{\text{St}}(\tau) \ket{v_1 v_2} +  (\rho_{ev,ge} + \rho_{ve,ge}) \bra{g_1 v_2} \hat{I}_{\text{St}}(\tau) \ket{v_1 v_2} \\
        &+ \rho_{eg,ge} \bra{g_1 v_2} \hat{I}_{\text{St}}(\tau) \ket{v_1 g_2} + \rho_{ge,eg} \bra{v_1 g_2} \hat{I}_{\text{St}}(\tau) \ket{g_1 v_2} + (\rho_{ev,ve } + \rho_{ve,ev})\bra{v_1 v_2} \hat{I}_{\text{St}}(\tau) \ket{v_1 v_2}.
    \end{split}
\end{equation}
Thus, $G_{coh,I}^{(2)}(\tau)$ and $ G_{coh,\rho}^{(2)}(\tau)$ depend on quantum coherence, while $ G_{d}^{(2)}(\tau)$ does not.

Notably, the contribution of $G_{d}^{(2)}(\tau)$ to the intensity correlation, which is $g_{\text{St,d}}^{(2)}(\tau)=G_{d}^{(2)}(\tau)/\braket{\hat{I}_{\text{St}(0)}}^2$, can be approximated by the Stokes-shifted correlation $\bar{g}_{\text{St}}^{(2)} (\tau)$ [Eq. (\ref{EqSM:g2_statistical_approach})] obtained within the conditional-probability approach described in Section SIV of this Supplemental Material. To show this, we take into account that, when the emitters couple only through the ZPL, the populations $p_{ev}$ and $p_{ve}$ in Eqs. (\ref{EqSM:mean_value_intensity_Stokes}) and (\ref{EqSM:Gd}) are very small in comparison to the populations of the purely electronic states, and the coherence $\rho_{ev,ve}$ vanishes. Neglecting these populations and coherence, the Stokes-shifted intensity can be approximated as $\braket{\hat{I}_{\text{St}(0)}}/ |\xi_{\text{St}}|^2\approx 2p_{ee}+p_{eg}+p_{ge}$. Further, as we consider that the vibrations decay very fast in comparison to the electronic states, we can additionally approximate the vibrational states by the electronic ground states in Eq. (\ref{EqSM:Gd}) [e.g., replacing $p_{ee}\bra{v_1 e_2}\hat{I}_{\text{St}}(\tau)\ket{v_1 e_2}$ by $p_{ee}\bra{g_1 e_2}\hat{I}_{\text{St}}(\tau)\ket{g_1 e_2}$]. Under these approximations, we can write $G_{d}^{(2)}(\tau) / |\xi_{\text{St}}|^4 \approx p_{ee}\bra{g_1 e_2}\hat{I}_{\text{St}}(\tau)\ket{g_1 e_2} + p_{ee} \bra{e_1 g_2}\hat{I}_{\text{St}}(\tau)\ket{e_1 g_2} + p_{eg} \bra{g_1 g_2}\hat{I}_{\text{St}}(\tau)\ket{g_1 g_2} + p_{ge} \bra{g_1 g_2}\hat{I}_{\text{St}}(\tau)\ket{g_1 g_2} $. As a result, we find $g_{\text{St,d}}^{(2)}(\tau)=G_{d}^{(2)}(\tau)/\braket{\hat{I}_{\text{St}(0)}}^2 \approx\bar{g}_{\text{St}}^{(2)}(\tau) $, which we have verified numerically.

In the following, we provide further insights of the three contributions in which we have decomposed $G_{\text{St}}^{(2)}(\tau)$ by deriving analytical expressions at $\tau=0$. With this aim, we first evaluate the expression of the operator $\hat{I}_{\text{St}}(\tau)$ in Eq. (\ref{EqSM:I_stokes}) at such time delay. On the one hand, we find that all the elements $\bra{a_1 b_2} \hat{I}_{\text{St}}(\tau) \ket{c_1 d_2}$ in Eq. (\ref{EqSM:G2_coh_rho}) vanish at $\tau=0$, which results in
\begin{equation} \label{EqSM:G2_coh_rho_delay_0}
    G_{coh,\rho}^{(2)}(0)=0.
\end{equation}
We have found numerically that $G_{coh,\rho}^{(2)}(\tau)$ can become nonzero and comparable to $G_{d}^{(2)}(\tau)$ at larger times, as discussed below and illustrated in Fig. \ref{FigureSM:G2_decomposition}. On the other hand, we find that $G_{d}^{(2)}(0)$ and $G_{coh,I}^{(2)}(0) $ are identical and given by
\begin{equation}  \label{EqSM:G2_coh_I_delay_0}
\begin{split} 
    G_{d}^{(2)}(0) /|\xi_{\text{St}}|^4  = G_{coh,I}^{(2)}(0)/|\xi_{\text{St}}|^4  &= 2 p_{ee} .
\end{split}
\end{equation}
However, we have found numerically that $G_{coh,I}^{(2)}(\tau)/|\xi_{\text{St}}|^4 $ decays to $0$ in the timescale of the vibrational lifetime, in contrast to $G_{d}^{(2)}(\tau)/|\xi_{\text{St}}|^4 $ that remains almost constant a larger time. Thus, $G_{\text{St}}^{(2)}(\tau)/|\xi_{\text{St}}|^4 $ is equal to $4\rho_{ee}$ at $\tau=0$ and decays to $\approx 2\rho_{ee}$ in the timescale of the vibrational lifetime. As a consequence, the normalized intensity correlation $g_{\text{St}}^{(2)}(\tau)$ also decays to half its initial value in the same timescale, which explains the narrow peak emerging at $\tau=0$ in those figures of the main text where $g_{\text{St}}^{(2)}(0)$ is significantly larger than $0$.

By substituting Eqs. (\ref{EqSM:mean_value_intensity_Stokes}), (\ref{EqSM:G2_coh_rho_delay_0}) and (\ref{EqSM:G2_coh_I_delay_0}) into Eq. (\ref{Eq:correlation_Stokes_first_analytical}), we obtain that the Stokes-shifted correlation at $\tau=0$ is given by
\begin{equation} \label{EqSM:Stokes_correlation_delay_0}
        g_{\text{St}}^{(2)}(0)=\frac{4p_{ee}}{\biggr[2p_{ee} + p_{eg} + p_{ge} \biggr]^2} .
\end{equation}
Here, we have not taken into account the steady-state populations $p_{ev}$ and $p_{ve}$, as well as the steady-state coherence $2\text{Re}\rho_{ev,ve}$, because we have verified numerically that they are negligible, as the emitters couple only through the ZPL. Further, the ZPL correlation at $\tau=0$ can be written as \cite{VivasViana_PRR, JuanDelgado}
\begin{equation} \label{EqSM:ZPL_correlation_delay_0}
    g_{ZPL}^{(2)}(0)=\frac{4p_{ee}}{\biggr[2p_{ee} + p_{eg} + p_{ge} + 2\text{Re}\rho_{eg,ge} \biggr]^2} ,
\end{equation}
where $\rho_{eg,ge}=\bra{eg}\hat{\rho}_{ss} \ket{ge}$ is the steady-state coherence between the pure electronic states $\ket{eg}$ and $\ket{ge}$. Equation (\ref{EqSM:ZPL_correlation_delay_0}) also does not take into account $p_{ev}$ and $p_{ve}$, as they are negligible. Comparing Eqs. (\ref{EqSM:Stokes_correlation_delay_0}) and (\ref{EqSM:ZPL_correlation_delay_0}), we find that the correlation of ZPL and of Stokes-shifted photons can be different at $\tau=0$ when $\text{Re}\rho_{eg,ge}$ is comparable to the value of the steady-state populations $p_{eg}$ and $p_{ge}$. For example, the latter occurs in Fig. 2b of the main text, where the laser is tuned to the transition frequency of the subradiant state. The value of the coherence $\text{Re}\rho_{eg,ge}$ is significant for this laser frequency and intensity because the excitation paths of the subradiant and superradiant states strongly interfere, as discussed in Ref. \cite{JuanDelgado}.

Last, to illustrate the behaviour of the different contributions in which we have decomposed $G_{\text{St}}^{(2)} (\tau)$, we plot in Fig. \ref{FigureSM:G2_decomposition} the dependence on $\tau$ of $G_{d}^{(2)} (\tau)/\braket{\hat{I}_{\text{St}} (0)}^2$ (solid green line), $G_{coh,I}^{(2)} (\tau)/\braket{\hat{I}_{\text{St}} (0)}^2$ (solid orange line), $G_{coh,\rho}^{(2)} (\tau)/\braket{\hat{I}_{\text{St}} (0)}^2$ (solid purple line) and of the total correlation $g_{\text{St}}^{(2)}(\tau)$ of Stokes shifted-photons (dashed red line). In particular, in Fig. \ref{FigureSM:G2_decomposition}a we use the same molecular configuration as in the simulations in Fig. 3a of the main text (i.e., the emitters are moderately coupled and driven by a laser tuned to the two-photon resonance $\omega_L = \omega_0$, with the rest of parameters detailed in the third row of Table \ref{Table}). We find that the two contributions to the numerator of the correlation of Stokes-shifted photons that involve quantum coherences vanish at large delay times, i.e., $G_{coh,I}^{(2)}(\tau\rightarrow\infty)=G_{coh,\rho}^{(2)}(\tau\rightarrow\infty)=0$. Additionally, $G_{coh,I}^{(2)}(\tau)$ is only non-zero at very short delay times, being responsible of the extremely narrow peak of the total correlation of Stokes-shifted photons $g_{\text{St}}^{(2)}(\tau)$ around $\tau=0$, as commented above. On the other hand, $G_{coh,\rho}^{(2)}(\tau)$ becomes important in the time-scale of the lifetime of the electronic excited states and can take either positive and negative values. A similar behaviour of these quantities is observed in Fig. \ref{FigureSM:G2_decomposition}b, where we fix the molecular configuration to that used in Fig. 3c of the main text, corresponding to two molecules separated by a much larger distance $r_{12}=400$ nm (the rest of parameters are specified in the fourth row of Table \ref{Table}).

\section{SVI. Experimental setup}

\begin{figure}
	\begin{center}
		\includegraphics[width=0.95\textwidth]{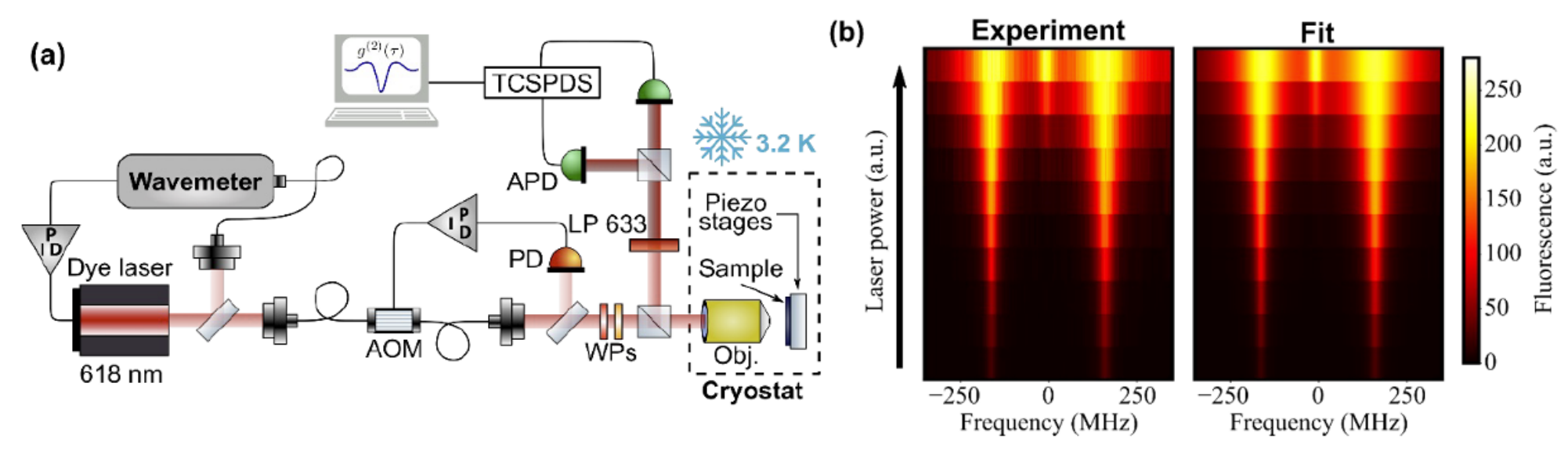}
		\caption{(a) Schematic representation of the experimental setup used to measure the correlation of Stokes-shifted photons $g_{\text{St}}^{(2)}(\tau)$. The intensity of the laser is stabilized using a photodiode (PD) in combination with an acousto-optic modulator (AOM). A wavemeter is employed to lock the laser frequency close to the molecular transition. The laser beam is tightly focused on two coupled emitters using a high-numerical-aperture microscope objective (NA 0.95), which also collects the emitted photons. A long-pass filter selectively transmits red-shifted photons above 633 nm, ensuring spectral isolation. The filtered photons are directed to the detection system, which comprises two avalanche photodiodes (APDs) connected to a time-correlated single-photon detection system (TCSPDS). (b) Experimental and fitted excitation spectra as a function of laser power (plotted using a logarithmic scale), ranging from 2 nW to 1024 nW. The experimental data points are compared to theoretical fits to extract the parameters of the pairs (coherent dipole-dipole coupling $V$, dissipative dipole-dipole coupling $\gamma_{12}$, and detuning $\delta$ between the two emitters).}  
        \label{FigureSM:setup}  
	\end{center}
\end{figure}

In this section we describe the experimental setup that we have used to measure the correlation of Stokes-shifted photons emitted from pairs of DBATT molecules. This setup is illustrated in Fig.~\ref{FigureSM:setup}a, where a continuous-wave (CW) laser (ring dye laser, Coherent 699) is employed to excite pairs of dibenzanthanthrene (DBATT) molecules, which are randomly embedded in a thin naphthalene film with a thickness of less than a few micrometers. To ensure stability, the laser frequency is digitally locked to a wavemeter (High Finesse WS7) with a relative precision of 1 MHz, preventing long-term drifts. The laser frequency can be manually adjusted by changing the lock point, allowing precise detuning with respect to the desired molecular transition. Additionally, the laser power is stabilized in a closed-loop system with a bandwidth of few kHz, using a fiber-coupled acousto-optic modulator (AOM). The AOM's transmission is controlled by adjusting its driving electrical power, ensuring consistent laser output for the experiments.

The sample is cooled to 3.2 K using a closed-cycle pulsed tube cryostat, where it is mounted on a piezo stage for precise positioning. The laser is focused using a high numerical aperture (NA = 0.95) microscope objective, which is also employed to collect emitted fluorescence photons. A long-pass filter (LP 633) is utilized to isolate Stokes-shifted photons while removing the ZPL and laser photons. The collected photons are directed to two avalanche photodiodes (SPCM-AQRH-15-FC, Excelitas) in a Start-Stop configuration to measure the intensity correlation. The time-histogram of photon detection statistics is reconstructed using a commercial Time-Correlated Single Photon Detection System (TCSPDS, TimeHarp 260), with a binning time width of $308$ ps. Due to the time resolution limitations of the avalanche photodiodes (APDs), which exhibit a jitter of approximately $400$ ps (FWHM), only frequencies below $1$ GHz can be reliably resolved in our measurements. As a result, our experimental setup is unable to capture the dynamics associated with the vibrational states, whose lifetimes are on the order of $10$ ps. This limitation justifies the absence of the extremely narrow peak at $\tau=0$ obtained in the simulated intensity correlation in Fig. 2c and Fig. 3 of the main text.

Moreover, the molecular detuning $\delta$, the coherent dipole-dipole coupling $V$ and the dissipative dipole-dipole 
coupling $\gamma_{12}$ of each pair are extracted through a two-dimensional fit of a series of excitation spectra measured at different laser intensities \cite{Trebbia_NatComms_2022}. The uncertainty in the values obtained of $V$, $\gamma_{12}$, and $\delta$ is $\approx 2 \%$. We have verified that this 2 \% uncertainty does not notably affect the simulated photon correlations. In the left panel of Fig. \ref{FigureSM:setup}b, we plot the experimental Stokes-shifted excitation spectra for the molecular pair presented in Figs. 3a and 3b of the main text. In these spectra, the transitions to the superradiant and subradiant states are clearly identified, as well as the central two-photon transition that becomes prominent when the laser power increases. Numerically, the excitation spectra is proportional to $\braket{\hat{I}_{\text{St}}(0)}$ [see Eq. (\ref{EqSM:mean_value_intensity_Stokes})]. The fit of $\braket{\hat{I}_{\text{St}}(0)}$ (which is shown in the right panel of Fig. \ref{FigureSM:setup}b and is in excellent agreement with the experimental excitation spectra) yields $\delta\approx5\gamma_0$, $V \approx 2.98\gamma_0$ and $\gamma_{12}\approx 0.29 \gamma_0$, with $\gamma_0=21.5$~MHz. Additionally, the transition dipole moments are found to be nearly parallel, forming a H-aggregate configuration (see Section SII of this Supplementary Material). We follow an identical procedure to fit the parameters of the molecular pair used in Fig. \ref{FigureSM:experiments} (which are specified in Table \ref{TableSM}), as well as to fit the parameters of the pair in Fig. 2 of the main text (see the second row of Table \ref{Table}, with experimental data reported in Ref. \cite{Trebbia_NatComms_2022}).

\section{SVII. Further experimental measurements}
\begin{table}[b]
    \centering
    \begin{tabular}{|c|c|c|c|c|c|c|c|c|c|c|c|c|c|}
        \hline
         Figure & $\boldsymbol{\mu}_j/|\boldsymbol{\mu}_j|$ & $\gamma_0 / (2\pi)$ [MHz] & $\alpha$ & $V/\gamma_0$ & $\gamma_{ij}/\gamma_0$ & $\delta/\gamma_0$ & $\Lambda/\gamma_0 $& $\omega_L$ & $\Omega_j/\gamma_0$ & $\boldsymbol{k}_L/|\boldsymbol{k}_L|$ & $\phi_{\text{St}}$ & $\hbar\omega_v$ [meV] & $1/\gamma_v$ [ps]\\
         \hline
         Fig. \ref{FigureSM:experiments}a      & $\hat{\boldsymbol{x}}$ & $21.5$ & $0.3$ & $9.19$ & $0.3$ & $41.44$ & $22.67$ & $\omega_0 + 21.95\gamma_0$ & $6.7$ & $\hat{\boldsymbol{y}}$ & $0$ & $31.86$ & $10$ \\
         \hline
         Fig. \ref{FigureSM:experiments}b &$\hat{\boldsymbol{x}}$ & $21.5$ & $0.3$ & $9.19$ & $0.3$ & $36.63$ & $20.49$& $\omega_0 - 6.98\gamma_0$& $7.44$ & $\hat{\boldsymbol{y}}$ & $0$ &$31.86$ & $10$ \\
        \hline
    \end{tabular}
    \caption{Molecular and laser parameters used in the simulations in Fig. \ref{FigureSM:experiments}a (second row) and in Fig. \ref{FigureSM:experiments}b (third row).}
    \label{TableSM}
\end{table}
\begin{figure}[t] 
	\begin{center}
		\includegraphics[width=0.95\textwidth]{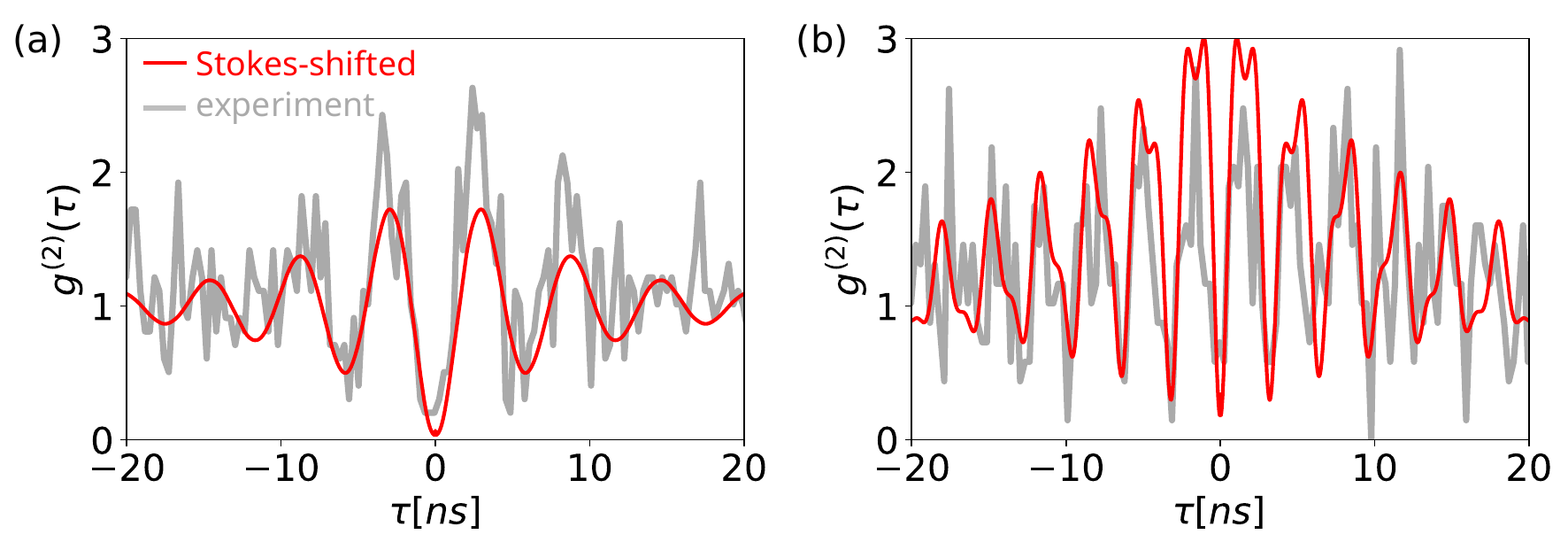}%
		\caption{Correlation of Stokes-shifted photons emitted from a pair of DBATT molecules in a H-aggregate configuration, with $V=9.19\gamma_0$ and $\gamma_{12}=0.3 \gamma_0$. Solid red lines correspond to the numerical result of the model that we propose in the main text, and solid grey lines correspond to experimental measurements. The laser frequency is $\omega_L = \omega_0 + 21.95 \gamma_0$ in (a), and $\omega_L = \omega_0 - 6.98 \gamma_0$ in (b). The rest of molecular and laser parameters are specified in Table \ref{TableSM}.}  
        \label{FigureSM:experiments}  
	\end{center}
\end{figure}
With the purpose of further verifying the accuracy of the model that we propose in the main text, we have performed additional experimental measurements of the correlation of Stokes-shifted photons emitted from a different pair of interacting DBATT molecules. This pair forms a molecular H-aggregate, with $V=9.19\gamma_0$ and dissipative coupling $\gamma_{12}\approx0.3 \gamma_0$, which is extracted by fitting the experimental excitation spectrum (as discussed in Section SVI of this Supplemental Material). In the experiment, we apply different voltages to the naphthalene crystal in which the molecules are immersed, which enables to modify the molecular detuning $\delta$ \cite{Trebbia_NatComms_2022}. 

First, in Fig. \ref{FigureSM:experiments}a we show the experimental measurements (grey line) obtained when the laser is slightly detuned from the superradiant state, similarly to Fig. 2a of the main text. However, in contrast to Fig. 2a of the main text, here the molecular detuning $\delta$ is considerably larger than the dipole-dipole coupling $V$ and, additionally, the molecular pair is driven by a more intense laser. We find that the intensity correlation resembles that in Fig. 2a, but exhibiting faster oscillations with the time delay. The experimental measurement is well reproduced by the simulated intensity correlation (red line), which is obtained using the complete model presented in the main text and considering again normal incidence and detection of light (the rest of parameters are specified in the second row of Table \ref{TableSM}).  

Moreover, in the experiment shown in Fig. \ref{FigureSM:experiments}b the molecular detuning is also large compared to the dipole-dipole coupling strength but, in contrast to Fig. \ref{FigureSM:experiments}a, the laser frequency is now moderately detuned from the two-photon resonance. In this case, the experimental intensity correlation exhibits even faster oscillations with the time delay, which are well reproduced by our simulations (all parameters used in the simulation are specified in the third row of Table \ref{TableSM}). 

\section{SVIII. Impact of a larger number of vibrational modes}
\begin{figure}[t] 
	\begin{center}
		\includegraphics[width=0.95\textwidth]{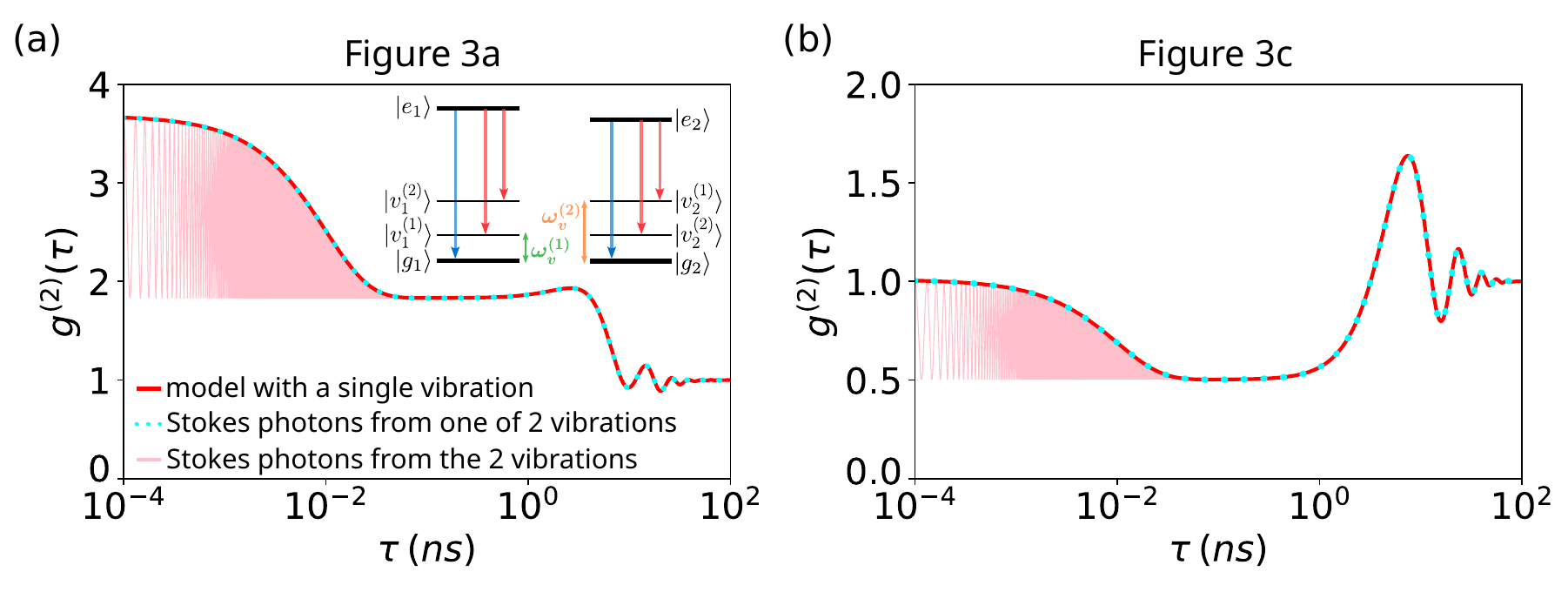}%
		\caption{Comparison of the simulated correlation of Stokes-shifted photons emitted from two DBATT molecules when different number of $1$-phonon states are considered. Red lines represent simulations obtained with the model presented in the main text, where a single $1$-phonon state is considered in each emitter, which corresponds to the vibrational mode at $257$ cm$^{-1}$ of the DBATT molecules. Additionally, we perform simulations using the Master Equation in Eq. (\ref{equation:master_equation_2vibrations}), where two $1$-phonon states are considered, $\ket{v_{j}^{(1)}}$ (with frequency $\omega_v^{(1)}$ corresponding again to the vibrational mode at $257$ cm$^{-1}$ of the DBATT molecules) and $\ket{v_{j}^{(2)}}$ (with frequency $\omega_v^{(2)}$ corresponding to the vibrational mode at $1331$ cm$^{-1}$ of the DBATT molecules). In the inset in (a) we depict the energy levels within this latter model, with the red arrows corresponding to the different Stokes-shifted transitions and the blue arrows to the ZPL transitions. We use the model that includes two $1$-phonon levels for each emitter to simulate the correlation of Stokes-shifted photons in two different scenarios: (i) considering the Stokes-shifted photons emitted due to the assistance of the two vibrational levels of each emitter (solid pink lines); and (ii) considering only the Stokes-shifted photons emitted due to the assistance of the vibrational mode of frequency $\omega_v^{(1)}$ (dotted cyan lines), which experimentally corresponds to filtering only the Stokes-shifted photons from this vibrational mode. In (a) we fix the molecular configuration to that used in Fig. 3a of the main text, where two molecules that interact strongly are driven by a laser tuned to the two-photon resonance $\omega_L = \omega_0$ (all the parameters are detailed in the third row of Table \ref{Table}), and in (b) we simulate the molecular configuration used in Fig. 3c of the main text, where the two emitters are at very far distances and again $\omega_L = \omega_0$ (all the parameters are specified in the fourth row of Table \ref{Table}).}   
        \label{FigureSM:2vibrations}  
	\end{center}
\end{figure}
In this section we analyze the impact of including additional $1$-phonon states in the Hamiltonian and the master equation, corresponding to other vibrational modes in the electronic ground state. More specifically, we consider that each emitter supports two vibrational modes of frequencies $\omega_v^{(1)}$ and $\omega_v^{(2)}$, with corresponding $1$-phonon states denoted as $\ket{v_j^{(n)}}$ (here, $j=1,2$ labels each emitter). We represent an scheme of the energy levels of the uncoupled emitters in the inset of Fig. \ref{FigureSM:2vibrations}a. The Hamiltonian of the system can thus be written as 
\begin{equation} \label{EqSM:Hamiltonian_2emitters_nvibrations}
\H = \hbar \sum_{j=1}^2 \biggr[ \frac{\Delta_j}{2} ( \ket{e_j}\bra{e_j}-\ket{g_j}\bra{g_j}) + \sum_{n=1}^2 \frac{2\omega_v^{(n)} - \Delta_j}{2}\ket{v_j^{(n)}}\bra{v_j^{(n)}} \biggr] + \H_I + \H_P,
\end{equation}
where the interaction and pumping Hamiltonians are again given as $\H_{I}=\hbar V(\sd_1 \s_2 + \s_1 \sd_2)$ and $\H_{P}=-\frac{\hbar}{2} \sum_{j=1}^2 (\Omega_j \sd_j + \Omega_j^* \s_j)$, respectively. The Markovian master equation including the dissipative processes and governing the dynamics of the system is given by
\begin{equation} \label{equation:master_equation_2vibrations}
    \frac{d}{dt}\dm =-\frac{i}{\hbar}[\H, \dm] + \sum_{j=1}^2 \biggr(\frac{\alpha \gamma_{{0}}}{2}\mathcal{D}[\s_j]+ \sum_{k\neq j} \frac{\gamma_{jk}}{2} \mathcal{D}[\s_j , \s_k]  + \sum_{n=1}^2 \biggr[ \frac{(1-\alpha) \gamma_{{0}}}{4}\mathcal{D}[\ket{v_j^{(n)}}\bra{e_j}]     + \frac{\gamma_{v}}{2}\mathcal{D}[\ket{g_j}\bra{v_j^{(n)}}] \biggr] \biggr)\rho , 
\end{equation}
where we have assumed that the decay rate from $\ket{e_j}$ to $\ket{v_{j}^{(1)}}$ is $(1-\alpha)\gamma_{{0}} /2$ and identical to the decay rate from $\ket{e_j}$ to $\ket{v_{j}^{(2)}}$. These latter dissipation processes are represented by red arrows in the inset in Fig. \ref{FigureSM:2vibrations}a, whereas the dissipation in the Zero-Phonon Line (having decay rate $\alpha \gamma_{{0}}$) is represented with blue arrows.  

Next we simulate again the correlation of Stokes-shifted photons emitted from two DBATT molecules at cryogenic temperatures. We use $\hbar \omega_v^{(1)}=31.86$ meV and $\hbar \omega_v^{(2)}=165.02$ meV, which correspond to the the $257$ cm$^{-1}$ and $1331$ cm$^{-1}$ vibrational modes, respectively, of the DBATT molecules  \cite{Jelezko_JCP_1997}. In Fig. \ref{FigureSM:2vibrations}a we fix the same molecular configuration as in Fig. 3a of the main text, where the two molecules strongly interact, and in Fig. \ref{FigureSM:2vibrations}b we fix the same molecular parameters as in Fig. 3c, where the two molecules are separated by a large distance (the rest of parameters are specified in the third and fourth rows of Table \ref{Table}, respectively). In both cases, the laser is tuned to the two-photon resonance $\omega_L = \omega_0$. We use $\E_{\text{St}}^{(+)} (t) / \xi_{\text{St}} = \sum_{n=1}^2 \ket{v_1^{(n)}}\bra{e_1} (t) + \ket{v_2^{(n)}}\bra{e_2} (t)$ to calculate the intensity correlation of Stokes-shifted photons emitted from the two molecules due to the assistance of the two vibrational modes. The resulting intensity correlation (solid pink lines) shows the same behaviour than the simulations obtained in the main text including a single vibrational mode in the master equation (red solid lines) at delay times larger than the lifetime of the vibrations ($\tau \gg 1/\gamma_v$). However, in the timescale of the vibrations lifetime we find that the decay of the envelope of the intensity correlation to half its value at $\tau=0$ (which is due to the loss of coherence, as discussed in the main text and in Section SV of this Supplemental Material) is accompanied by fast oscillations of frequency $\omega_v^{(2)}-\omega_v^{(1)}$ due to the inclusion of two vibrations. Additionally, we calculate the intensity correlation using $\E_{\text{St}}^{(+)} (t) / \xi_{\text{St}} =  \ket{v_1^{(1)}}\bra{e_1} (t) + \ket{v_2^{(1)}}\bra{e_2} (t)$ (cyan dotted lines), which represents the correlation of Stokes-shifted photons that are due only to the assistance of the vibrational mode at $257$ cm$^{-1}$. We find that this latter simulated intensity correlation is identical to the one obtained in the main text using a Markovian master equation that includes a single vibrational mode.

\section{SIX. Generalization of the model to $N$ emitters}
In this section, we show that the model can be generalized to describe the Zero-Phonon-Line and Stokes-shifted emissions from of $N$ almost identical emitters. Additionally, we consider that the emitter $j$ supports an arbitrary number $M_j$ of vibrational/phononic modes. In this case, the total Hamiltonian (in the rotating frame at the laser frequency) is given as
\begin{equation}  \label{EqSM:total_Hamiltonian_N_emitters}
\H = \hbar \sum_{j=1}^N \biggr[ \frac{\Delta_j}{2} ( \ket{e_j}\bra{e_j}-\ket{g_j}\bra{g_j}) + \sum_{n}^{M_j} \frac{2\omega_{v}^{(j,n)} - \Delta_j}{2}\ket{v_j^{(n)}}\bra{v_j^{(n)}} \biggr] + \H_I + \H_P ,
\end{equation}
where $\omega_{v}^{(j,n)}$ is the frequency of the $n^{\text{th}}$ vibrational/phononic mode of the emitter $j$. The pumping Hamiltonian can be written as
\begin{equation}
\H_{P}=-\frac{\hbar}{2} \sum_{j=1}^N (\Omega_j \sd_j + \Omega_j^* \s_j) .
\end{equation}
Additionally, the interaction Hamiltonian $\H_I$ in Eq. (\ref{EqSM:total_Hamiltonian_N_emitters}) includes all possible dipole-dipole couplings between different emitters. Specifically,
\begin{equation}
    \H_{I}=\hbar \sum_i^N \sum_{j<i}^N V_{ij} (\sd_i \s_j + \s_i \sd_j) ,
\end{equation}
where
\begin{equation} \label{EqSM:V_ij}
        V_{ij} = \frac{3}{4} \alpha \gamma_{0} \biggr[- \biggr(\hat{ \boldsymbol{\mu}}_i \cdot\hat{ \boldsymbol{\mu}}_j - (\hat{ \boldsymbol{\mu}}_i \cdot \hat{\boldsymbol{r}}_{ij})(\hat{ \boldsymbol{\mu}}_j \cdot \hat{\boldsymbol{r}}_{ij}) \biggr)\frac{\cos(k_{ij} r_{ij})}{k_{ij} r_{ij}} + \biggr(\hat{ \boldsymbol{\mu}}_i \cdot \hat{ \boldsymbol{\mu}}_j - 3(\hat{ \boldsymbol{\mu}}_i \cdot \hat{\boldsymbol{r}}_{ij})(\hat{ \boldsymbol{\mu}}_j \cdot \hat{\boldsymbol{r}}_{ij}) \biggr)\biggr(\frac{\sin(k_{ij} r_{ij})}{(k_{ij} r_{ij})^2} + \frac{\cos(k_{ij} r_{ij})}{(k_{ij} r_{ij})^3}\biggr) \biggr]
\end{equation}
is the dipole-dipole coupling strength between emitters $i$ and $j$, which have transition dipole moments $\boldsymbol{\mu}_i $ and $\boldsymbol{\mu}_j$, and are located at positions $\boldsymbol{r}_i$ and $\boldsymbol{r}_j$. In Eq. (\ref{EqSM:V_ij}), we have also introduced $\boldsymbol{r}_{i}-\boldsymbol{r}_{j}=r_{ij}\hat{\boldsymbol{r}}_{ij}$, with $r_{ij}$ the distance between the emitters $i$ and $j$. Additionally, $\gamma_0$ is again the total decay rate of the excited state, which is considered identical for all emitters, and $k_{ij}$ is the wavenumber of the arithmetic mean of the transition frequencies of the emitters $i$ and $j$.

Moreover, the dissipators in the Markovian master equation in Eq. (\ref{equation:master_equation_2vibrations}) can be directly generalized as
\begin{equation} \label{equation:master_equation}
    \frac{d}{dt}\dm =-\frac{i}{\hbar}[\H, \dm] + \sum_{j=1}^N \biggr(\frac{\alpha \gamma_{{0}}}{2}\mathcal{D}[\s_j]+ \sum_{k\neq j} \frac{\gamma_{jk}}{2} \mathcal{D}[\s_j , \s_k]  + \sum_{n=1}^{M_j} \biggr[ \frac{\gamma_{e-v}^{(j,n)}}{2}\mathcal{D}[\ket{v_j^{(n)}}\bra{e_j}]     + \frac{\gamma_{v}^{(j,n)}}{2}\mathcal{D}[\ket{g_j}\bra{v_j^{(n)}}] \biggr] \biggr)\hat\rho , 
\end{equation}
where the dissipative dipole-dipole coupling between emitters $i$ and $j$ is given by
\begin{equation}
    \gamma_{ij} = \frac{3}{2} \alpha \gamma_{0}  \biggr[ \biggr(\hat{ \boldsymbol{\mu}}_i \cdot \hat{ \boldsymbol{\mu}}_j - (\hat{ \boldsymbol{\mu}}_i \cdot \hat{\boldsymbol{r}}_{ij})(\hat{ \boldsymbol{\mu}}_j \cdot \hat{\boldsymbol{r}}_{ij}) \biggr)\frac{\sin(k_{ij} r_{ij})}{k_{ij} r_{ij}} + \biggr(\hat{ \boldsymbol{\mu}}_i \cdot \hat{ \boldsymbol{\mu}}_j - 3(\hat{ \boldsymbol{\mu}}_i \cdot \hat{\boldsymbol{r}}_{ij})(\hat{ \boldsymbol{\mu}}_j \cdot \hat{\boldsymbol{r}}_{ij}) \biggr)\biggr(\frac{\cos(k_{ij} r_{ij})}{(k_{ij} r_{ij})^2} - \frac{\sin(k_{ij} r_{ij})}{(k_{ij} r_{ij})^3}\biggr) \biggr] .
\end{equation}
$\gamma_{v}^{(j,n)}$ in Eq. (\ref{equation:master_equation}) is the decay rate of the $n^{\text{th}}$ vibrational/phononic mode of the emitter $j$ (i.e., $1/\gamma_{v}^{(j,n)}$ is the lifetime of this vibrational/phononic mode), while $\gamma_{e-v}^{(j,n)}$ is the decay rate of the excited state $\ket{e_j}$ into the $1$-phonon level $\ket{v_j^{(n)}}$. Notably, as the total decay rate of emitter $j$ is $\gamma_{{0}}$ and the decay rate in the ZPL is $\alpha \gamma_{{0}}$, the set of decay rates into of $1$-phonon levels satisfy
\begin{equation}
 \sum_n^{M_j} \gamma_{e-v}^{(j,n)} = (1-\alpha) \gamma_{{0}} .
\end{equation}

The correlation of ZPL photons can be obtained by substituting the positive-frequency electric field operator $\hat{E}_{ZPL}^{(+)}(t)/\xi_{ZPL} = \sum_{j=1}^N \ket{g_j}\bra{e_j} (t)$ into the expression for $g^{(2)}(\tau)$ in Eq. (3) of the main text, which takes into account that the emitters have identical transition dipole moments. Additionally, the correlation of all Stokes-shifted photons can be obtained through the electric field operators \cite{Loudon_book_2000}
\begin{equation} \label{EqSM:general_Stokes_field}
    \hat{E}_{\text{St}}^{(+)}(t)/\tilde{\xi}_{\text{St}} = \sum_{j=1}^N \sum_{n=1}^{M_j} \sqrt{\gamma_{e-v}^{(j,n)}  \biggr(\omega_0 - \omega_{v}^{(j,n)} \biggr)} \ket{v_j^{(n)}}\bra{e_j} (t) ,
\end{equation}
with $\tilde{\xi}_{\text{St}}$ a new proportionality constant that only depend on the distance to the detectors and physical constants, and which does not affect the calculation of the intensity correlation. We note that the square root factor in Eq. (\ref{EqSM:general_Stokes_field}) has been neglected in Section SVIII for simplicity, as the value of this factor is almost identical for the two molecular vibrations considered there.


\bibliography{bibliography.bib}